\documentclass[acs,preprint,preprintnumbers,floatfix,amsmath,amssymb]{revtex4}
\usepackage{xspace}
\usepackage{psfrag,graphicx} 

\begin{document}

\newcommand\eurpje{Eur.~Phys.~Journ.~E}
\newcommand\jpca{J.~Phys.~Chem.~A}
\newcommand\biopj{Biophys.~J.}
\newcommand\aciee{Angew.~Chem.~Intl.~Ed.~Eng.}
\newcommand\tetra{Tetrahedron}
\newcommand\cossms{Curr.~Op.~Sol.~State and Mat.~Sci.}
\newcommand\jpb{J.~Phys.~B}
\newcommand\pnas{Proc.~Natl.~Acad.~Sci~USA.}
\newcommand\cpl{Chem.~Phys.~Lett.}
\newcommand\sci{Science}
\newcommand\jcop{J.~Comp.~Phys.}
\newcommand\rpphys{Rep.~Prog.~Phys.}
\newcommand\philtrans{Phil.~Trans.~Roy.~Soc.~A}
\newcommand\jpcb{J.~Phys.~Chem.~B}
\newcommand\ptrsa{Phil.~Trans.~Roy.~Soc.~A}
\newcommand\jmb{J.~Med.~Biol.}
\newcommand\jacs{J.~Am.~Chem.~Soc.}
\newcommand\pccp{Phys.~Chem.~Chem.~Phys.}

\title{Design principles for Bernal spirals and helices with tunable pitch}

\author{Szilard N.~Fejer$^{a,b}$}
\email{szilard.fejer@cantab.net}
\author{Dwaipayan~Chakrabarti$^{c}$}
\email{d.chakrabarti@bham.ac.uk}
\author{Halim~Kusumaatmaja$^{d}$}
\email{halim.kusumaatmaja@durham.ac.uk}
\author{David J.~Wales$^{e}$}
\email{dw34@cam.ac.uk}
\affiliation{
$^a$ Department of Chemical Informatics, University of Szeged, Faculty of Education,
Boldogasszony sgt.~6, H-6725 Szeged, Hungary \\
$^b$ Pro-Vitam Ltd., str.~Muncitorilor nr.~16, 520032 Sfantu Gheorghe, Romania \\
$^c$ School of Chemistry, University of Birmingham, Edgbaston, Birmingham B15 2TT, United Kingdom\\
$^d$ Department of Physics, Durham University, South Road, Durham DH1 3LE, United Kingdom\\
$^e$ University Chemical Laboratories, Lensfield Road, Cambridge CB2 1EW, United Kingdom
}

\begin{abstract}
Using the framework of potential energy landscape theory, we describe two {\it in silico} designs 
for self-assembling helical colloidal superstructures based upon dipolar dumbbells and 
Janus-type building blocks, respectively. Helical superstructures with controllable pitch length are obtained
using external magnetic field driven assembly of asymmetric dumbbells involving screened electrostatic 
as well as magnetic dipolar interactions. The pitch of the helix is tuned by modulating the Debye screening 
length over an experimentally accessible range. The second design is based on building blocks 
composed of rigidly linked spheres with short-range anisotropic interactions, which are predicted 
to self-assemble into Bernal spirals. These spirals are quite 
flexible, and longer helices undergo rearrangements {\it via} cooperative, hinge-like moves, in 
agreement with experiment. 

{\bf keywords:} self-assembly, chirality, Bernal spirals, Janus particles, mesoscopic modelling
\end{abstract}

\maketitle

\section{Introduction}

The ubiquitous presence of helical architectures in nature, as well as their
diverse potential applications in materials science, for
optoelectronics \cite{PercecETAL02}, sensors \cite{YashimaMIFN09}, responsive
materials \cite{PijperF08}, and asymmetric catalysis \cite{GierBFS98}, has
motivated interest in design and synthesis.
Molecular self-assembly, in particular,
is a promising route to helicity \cite{Lehn02,YashimaMIFN09}. Self- or
directed-assembly of nanoparticles and colloidal building blocks has enormous
potential as a means of fabrication because of the scope for tuning
the interactions \cite{GlotzerS07,SacannaICP10}. A delicate balance between a
variety of weak forces often governs the assembled
structure \cite{MinAKGI08}. A thorough understanding of these forces holds the
key to rational design. 

The present contribution reports on two complementary routes to helical
nanostructures, starting from anisotropic building blocks.
The first strategy employs directed assembly of achiral
colloidal building blocks \cite{ChakrabartiFW09,ZerroukiBPCB08}, where an
interplay between two length scales for the anisotropic interactions determines
the emergent chirality of the nanostructure. Such competing length scales are
present in DNA \cite{WatsonC53}, one characterising the distance between
consecutive nucleotides in the sugar-phosphate backbone, and the other
governing the stacking of the base pairs. Here the
competing length scales arise due to electrostatic and
magnetic dipolar interactions. While much progress has been made in obtaining emergent
chirality from achiral building blocks, biasing the superstructure to a
particular handedness \cite{HowsonETAL12}, or controlling the
pitch \cite{SrivastavaETAL10}, has proved more difficult. We address the latter
challenge using theory and simulation, and demonstrate that significant control
(around $30\%$) over the pitch length can be achieved by modulating the Debye screening
length of the electrostatic interactions over an experimentally accessible
range. The resulting tunable pitch length for helical superstructures 
holds significant promise for 
the design of a novel class of responsive materials.

The second design principle considered here involves clusters
of Janus particles. Several models have been used recently to study
the dynamics and aggregation properties of systems composed of Janus-type building
blocks \cite{janus_langmuir_2008,janus_softmatter_2012,janus_langmuir_2012,janus_langmuir_2013},
and the resulting phase diagrams exhibit a wide variety of potential target morphologies
for self-assembly, depending on the anisotropic
properties (shape and interactions). Interestingly, none of the models employed
so far was able to reproduce assembly into Bernal spirals (BC spirals or
tetrahelices) from anisotropic building blocks, although such systems have been designed and observed
experimentally \cite{janus_2011}. Here we report the design of a Janus building
block that prefers assembly into Bernal spirals, suitably guided by
experimentally relevant, anisotropic interaction potentials \cite{janus_langmuir_2008,janus_2011}.

Assembly into one-dimensional polytetrahedral clusters, locally organized as Bernal spirals, has also been achieved 
for isotropic particles \cite{Campbell2005}, and reproduced computationally, by tuning the balance 
between a long-range screened isotropic repulsion 
and a short-range attraction term in the potential \cite{Mossa2004,Sciortino2005}.

\section{Methods}

We have used basin-hopping global optimisation \cite{Scheraga87,WalesD97,waless99} as implemented in the {\tt GMIN} program \cite{GMIN} 
to predict global minima. Basin-hopping global optimisation involves perturbations of geometry
followed by energy minimisation.
The perturbations are designed to avoid any overlapping
particles. For the dipolar dumbbells, we run 50000 basin-hopping steps for each set of parameters presented in this paper.
Global minima for Janus clusters have been identified by running 10000 basin-hopping steps for at least 10 random starting 
structures.

The energy landscapes for 20 and 24 Janus particles described in \S \ref{sec:Janus} were explored
using double-ended pathway searches between local minima with the discrete path sampling \cite{Wales02,Wales04,Wales06} approach,
as implemented in our {\tt OPTIM} \cite{optim} and {\tt PATHSAMPLE} \cite{pathsample} programs. 
We have employed the doubly-nudged \cite{trygubenkow04} elastic band \cite{henkelmanj99,HenkelmanUJ00,HenkelmanJ00} 
(DNEB) method \cite{trygubenko04} to locate transition state 
candidates. The method has been adapted to avoid overlapping geometries for 
Janus building blocks, by diagnosing overlap between the ellipsoids in each interpolated structure
and moving the overlapping ellipsoids by small random amounts until there is no overlap in the cluster.
Transition states were refined using gradient-only hybrid eigenvector-following \cite{MunroW99} from TS candidates
identified with the DNEB algorithm.
The most likely rearrangement mechanisms (pathways with the largest contribution to 
the steady-state rate constant ignoring recrossings \cite{Wales06}) were obtained using Dijkstra analysis \cite{Dijkstra59}
of the resulting kinetic transition network.

To visualise the corresponding
multidimensional potential energy surfaces we 
construct disconnectivity 
graphs \cite{BeckerK97,WalesMW98} from the databases of minima and transition states explored during discrete path sampling.
Further details of all the geometry optimisation and visualisation techniques exploited
in this potential energy landscapes framework can be found in previous
reports and reviews \cite{Wales03,Wales06,Wales10a,Wales12}.

\section{Controllable Helix Pitch}

\subsection{A Decorated Rigid Body Model}
The colloidal building blocks considered here are charged dipolar asymmetric dumbbells, which 
involve screened electrostatic as well as dipolar interactions. We modelled these particles 
using multiple interaction sites that decorate a rigid framework. Each dumbbell involves two 
lobes, each modelled by a spherically symmetric effective Yukawa pair potential \cite{RobbinsKG92,LowenK93},
where the inverse screening length $\kappa$ controls the range as well as the softness of the 
screened electrostatic interactions, which can be tuned in experiment by modulating the salt 
concentration of the medium \cite{LowenK93,YethirajB03}. Additionally, there is a magnetic point 
dipole between the lobes, directed perpendicular to the axis. 
The total energy of a system of $N$ such dumbbells in an external magnetic field $\boldsymbol{B}$ is given by
\begin{eqnarray}
U &=& \displaystyle \sum_{I=1}^{N-1} \displaystyle \sum_{J=I+1}^{N}
\displaystyle \sum_{i\in I}^{1,2} \displaystyle \sum_{j \in J}^{1,2}
\epsilon_{ij} \frac{ \exp [ - \kappa(r_{ij} - \sigma_{ij})]}{r_{ij}/\sigma_{ij}} \nonumber \\ 
&+& \sum_{I=1}^{N-1} \sum_{J=I+1}^{N} \frac{\mu_{D}^{2}}{r_{IJ}^{3}}
\Biggl[(\boldsymbol{\hat \mu}_{I} \cdot \boldsymbol{\hat \mu}_{J}) - 3 (\boldsymbol{\hat \mu}_{I} \cdot
{\bf \hat r}_{IJ}) (\boldsymbol{\hat \mu}_{J} \cdot \boldsymbol{\hat r}_{IJ}) \Biggr] 
- \mu_{D}\displaystyle \sum_{I=1}^{N} \boldsymbol{\hat \mu}_{I} \cdot \boldsymbol{B}.
\label{eq:dmbl}
\end{eqnarray}
Here, ${\bf r}_{I}$ is the position vector for the magnetic point dipole on dumbbell
$I$, $\boldsymbol{\hat \mu}_{I}$ is the unit vector defining the direction of
the dipole moment, whose magnitude is $\mu_{D}$, ${\bf r}_{IJ} = {\bf r}_{I} -
{\bf r}_{J}$ is the separation vector between dipoles on dumbbells $I$ and $J$
with magnitude $r_{IJ}$, ${\bf \hat r}_{IJ} = {\bf r}_{IJ} / r_{IJ} $, and
$r_{ij}$ is the separation between the Yukawa sites $i$ and $j$. The units of
energy and length are chosen as the Yukawa parameters $\epsilon_{Y}$ and
$\sigma_{11}$, respectively. For the Yukawa interactions we set $\epsilon_{11}
= \epsilon_{22} = \epsilon_{12} = 0.1 \epsilon_{\rm Y}$ and $\sigma_{11} = 1$. 
$\sigma_{22} < 1$ defines the asymmetry parameter $\alpha = \sigma_{11}/\sigma_{22}$;
$\sigma_{12}$ is chosen to be the arithmetic average $\sigma_{12} =
(\sigma_{11} + \sigma_{22})/2$. The direction of the external field
$\boldsymbol{B} = (0, 0, B)$ was held fixed along the $z$-axis of the
space-fixed frame as its strength, $B$, was varied. The magnetic dipole
$\mu_{D}$ is then in reduced units of $(4\pi\epsilon_{Y}\sigma_{11}^3/\mu_0)^{1/2}$ and the magnetic
field strength $B$ is in $[\epsilon_{Y}\mu_{0}/(4\pi\sigma_{11}^3)]^{1/2}$, where
$\mu_{0}$ is the permeability of free space. For the simulation results presented here,
we have used in reduced units $\mu_{D} = 0.1$, $B = 10.0$, $\sigma_{11} = 1.0$, $\sigma_{22} = 0.4$. 
$\kappa$ is varied over an experimentally relevant range, as discussed in the next section.

\vspace{0.5cm}
\begin{figure}[h]
\centerline{\includegraphics[width=12cm]{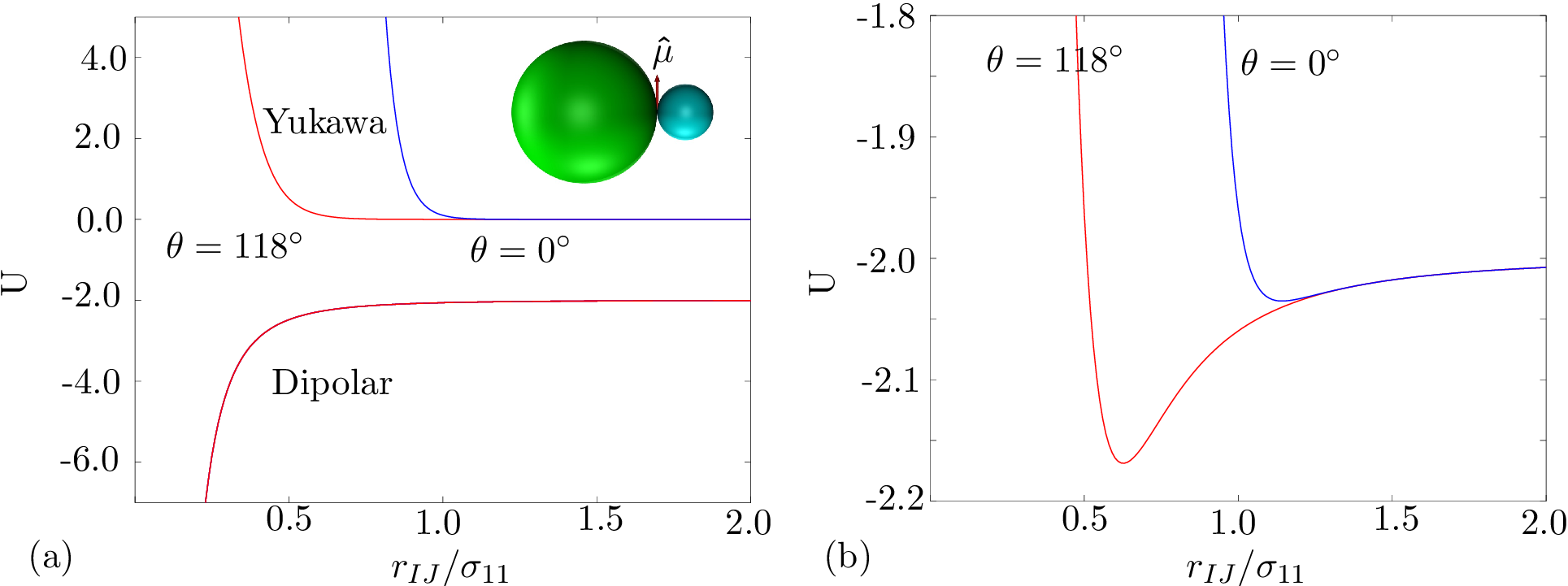}}
\caption{The potential energy of two interacting dumbbells with $\kappa
\sigma_{11} = 20$ as a function of the separation between dumbbells $I$ and $J$,
$r_{IJ}$.  The other parameters are as given in the text. In panel (a), the
contributions from the Yukawa and dipolar interactions are shown separately.
Panel (b) shows the sum of the two contributions. In the
figures, we have assumed that the direction of the dipole moment is parallel to
the external field. $\theta$ is the angle
between the axes of dumbbells $I$ and $J$; the global minimum energy for 
two dumbbells is located at $\theta \sim 118^\circ$. The inset in
panel (a) shows a schematic representation of the dumbbell building block,
where the point dipole perpendicular to the axis of the dumbbell is not drawn
to scale. }
\label{fig:hlxpitch}
\end{figure}

\subsection{Results}

For small clusters, the global optimisation results reveal a helical superstructure as the ground 
state for an optimal asymmetry when a sufficiently strong external field is applied. The dumbbells 
tend to align perpendicular to the field, to facilitate alignment of the dipoles.
An optimal asymmetry is critical for helix formation, because competition with a second length 
scale, which controls the steric interactions, is the basis of the emergent chirality \cite{ChakrabartiFW09,ZerroukiBPCB08}. 
A single helical strand is observed without any predetermined chirality for clusters up to at least 
$N = 20$ for the set of parameters investigated. 

\begin{figure}[h]
\centerline{\includegraphics[width=12cm]{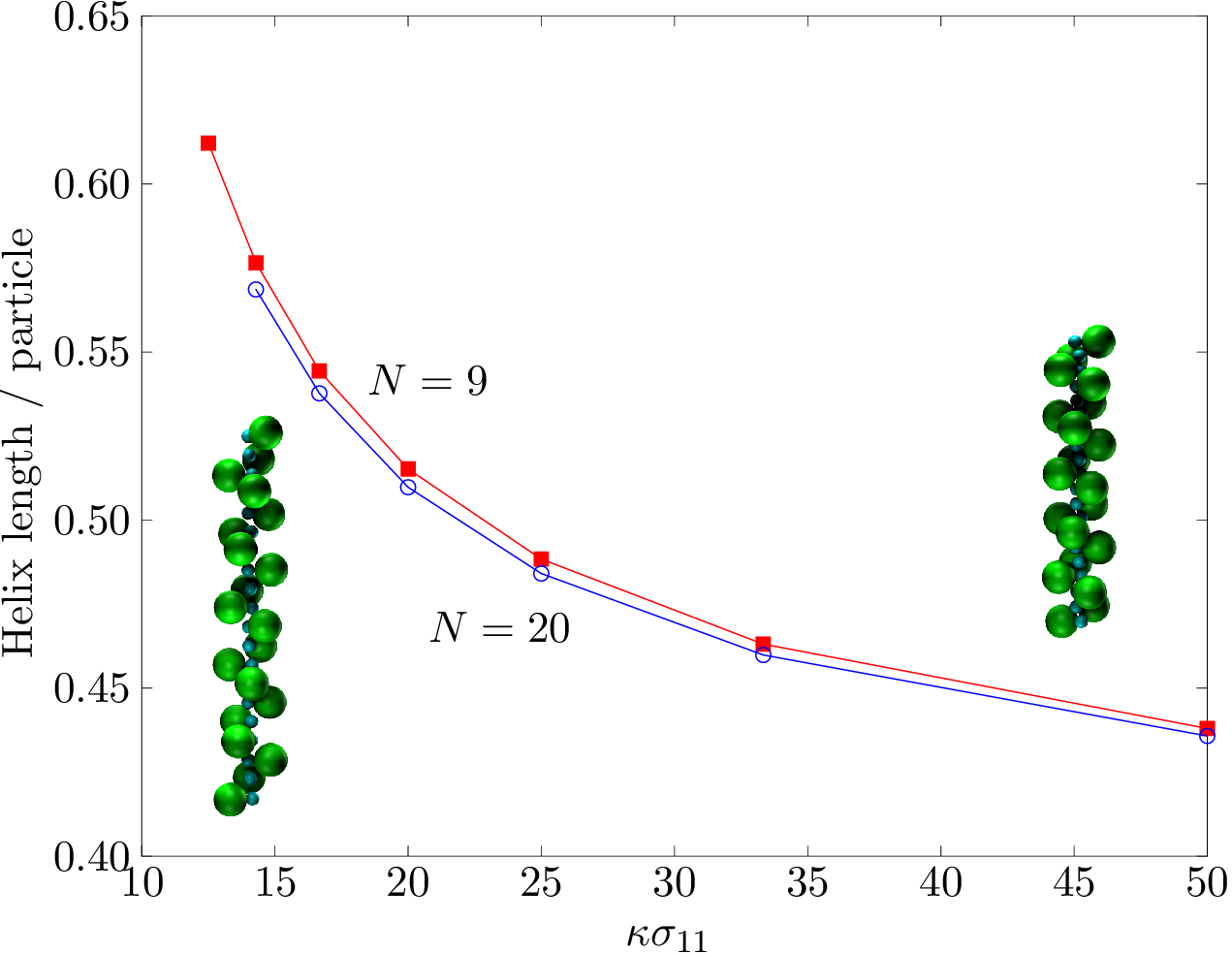}}
\caption{Helix length per particle as a function of the inverse screening length 
for the predicted ground state structures. Results are shown for $N = 9$ and $N=20$.
Insets: the ($N=20$) helix configurations for (left) $\kappa\sigma_{11} = 14.3$ and
(right) $\kappa\sigma_{11} = 50.0$. The dipoles are not shown for clarity of presentation.}
\label{fig:hlxpitch2}
\end{figure}

Figure \ref{fig:hlxpitch} shows that the pitch length can be controlled by modulating the range of 
the screened electrostatic interactions, which can be tuned experimentally by changing the salt 
concentration of the medium \cite{LowenETAL05}. It is evident that the pitch of the helix changes by 
nearly $30\%$ upon varying $\kappa$ over a range accessible in experiments for both $N = 9$ and 
$N = 20$. The slight difference between the two sizes arises due to additional (attractive) dipolar 
interactions when more dumbbells are present in the cluster. As $\kappa$ is increased (screening 
length is decreased), the Yukawa potential is shorter in range and the equilibrium distance between 
two dumbbells decreases. The change in pitch is primarily attributed to this varying equilibrium
separation, the change in twist angle being nominal. The limiting cases are insightful. For large 
$\kappa$, the Yukawa potential approaches a hard-sphere interaction, and for $\kappa \to 0$, it 
approaches the long-range Coulomb potential. Figure \ref{fig:hlxpitch2} shows that the range of the 
screened electrostatic interactions directly affects the helix pitch length, but helix integrity is 
preserved. Hence the design proposed here offers a route to helical nanostructures with controllable 
pitch length.

As for the parameters, a reasonable estimate in physically relevant units can be obtained by setting 
$\epsilon_{\rm Y} = 4.1 \times 10^{-21} \rm J$ (of the order of $k_{B}T$) and $\sigma_{11} = 10^{-6} \rm m$.
In the above analysis we have neglected the screening effect on the dipolar interactions in 
the medium. This assumption is valid when magnetic dipoles are involved \cite{vanBlaaderenETAL13,ZerroukiBPCB08}. 
In the absence of the screening effect for the magnetic interactions, the screened electrostatic and 
magnetic interactions can be manipulated independently \cite{SmoukovGMV09}on The values in reduced units 
used here correspond to a magnetic dipole moment $\mu_{D} \sim 2 \times 10^{-17}\,{\rm A\,m^{2}}$ and 
a magnetic field strength $B  \sim 2 \times 10^{-4}\,{\rm T}$, well within the 
experimentally accessible regime. 


If we consider an aqueous medium for a monovalent electrolyte, 
where the ionic strength $I$ is equal to the molar concentration, the Bjerumm length  
$\lambda_{\rm B}$ of water is $0.7\,{\rm nm}$ at $298\,{\rm K}$, equivalent to a colloid charge 
of $Z \sim 130$ for the larger lobe of the dumbbell when $\kappa\sigma_{11} = 20$ using the relationship
$\epsilon_{11}/(k_{\rm B}T) = Z^{2}(\lambda_{\rm B}/\sigma_{11})/(1+\kappa\sigma/2)^{2}$ \cite{IvlevLMR12}.
For the parameter range considered here, 
the variation of the Debye screening length $\kappa^{-1}$ is between $20\,{\rm nm}$ and $80\,{\rm nm}$. 
With concentrations as low as $\sim 1\,\mu M$ achievable in an experimental setup \cite{YethirajB03},
this range is well within the regime accessible in experiments, since the following relationship 
holds: $\kappa^{-1} = 0.304\,I^{-1/2}$, where $\kappa^{-1}$ is in nanometres and $I$ is in moles per 
litre \cite{Israelachvili11}. 

\vspace{0.5cm}

\section{Clusters of Janus Particles}
\label{sec:Janus}

\subsection{Computational Model}

We have previously shown that very different interparticle potentials can 
produce rather
similar preferred arrangements during aggregation \cite{fejer_self-assembly_2011}, 
as long as the overall pair potentials are sufficiently alike.
Current designs of spherical mesoscopic Janus building blocks generally involve a charged
hydrophilic hemisphere combined with a hydrophobic hemisphere
in an aqueous environment. Changing the ionic strength of the solution effectively changes the screening of the electrostatic repulsion,
and worm-like structures arise when the charges are well screened, 
corresponding to short-ranged repulsive terms.
Potentials that have been used for Janus particle modelling therefore usually include three types of interaction: 
hard sphere repulsion to prevent 
overlap, hydrophobic interactions, and screened Coulombic repulsion \cite{janus_langmuir_2008}.
Most potentials employed to date are discontinuous, containing quasi-square well functions and
hard sphere interactions, which are not suitable for energy landscape studies
based on geometry optimisation. Recently, a 
potential has been developed for soft Janus particles \cite{janus_softmatter_2012}, 
but it involves considerably longer-range
interactions than we consider in the present study.

Our design for a Janus building block tries to capture the net behaviour of
particles with strongly anisotropic short-range interactions in solution, aggregating around the 
hydrophobic hemisphere. We use continuous and differentiable functional forms, and aim to keep the potential as simple as possible,
to extract the minimal conditions on the interparticle forces that correspond to particular target morphologies.
The Paramonov-Yaliraki (PY) potential \cite{paramonov05} has proved its
versatility for modelling a large number of anisotropic interactions \cite{fejer_self-assembly_2011,FejerCW10,ChakrabartiFW09},
and here we have used this 
representation to create Janus-type particles by modifying just two interaction parameters in the pairwise energy.
Each Janus building block is composed of two rigidly linked spheres (A and B) represented by PY ellipsoids \cite{paramonov05,olesen_2013} 
having the same orientation and shifted along the $z$ axis by 0.1 distance units from the origin, in opposite directions. 
The building blocks interact within a rigid-body framework 
using the angle-axis description for the orientational degrees of freedom \cite{Wales05,ChakrabartiW09}.
To allow for shorter-range interactions than the usual Lennard-Jones form, 
we have increased the diameter of the spheres threefold, while keeping the range parameter $\sigma_0$ fixed at unity.
One sphere (ellipsoid A) is purely repulsive, while the other has a 
higher interaction strength along the $z$ direction (attractive
semiaxis length $a_{23}=1.56$). The total interaction energy between building blocks is
\begin{align}
U_{12}=4\epsilon_{0}\sum_{i=1}^{2}\sum_{j=1}^{2}\left[\epsilon_{\rm rep,i}\epsilon_{\rm rep,j}\left(\frac{\sigma_{0}}
{r_{ij}-r_{ij}F_{1ij}^{-1/2}+\sigma_{0}}\right)^{12}\right.-\nonumber\\
\left.-\epsilon_{\rm attr,i}\epsilon_{\rm attr,j}\left(\frac{\sigma_{0}}
{r_{ij}-r_{ij}F_{2ij}^{-1/2}+\sigma_{0}}\right)^{6}\right],
\label{eq:paramonov}
\end{align}
where $F_{1ij}$ and $F_{2ij}$ are the 
\lq repulsive' and \lq attractive' elliptic contact functions \cite{paramonov05}, 
calculated between ellipsoids $i$ and $j$, $\epsilon_{\rm rep}=1$
for both ellipsoids in the building block, and $\epsilon_{\rm attr,A}=0$, and $\epsilon_{\rm attr,B}=1$. The repulsive 
semiaxes for both ellipsoids are $a_{11}=a_{12}=a_{13}=1.5$. The attractive semiaxes are not used for ellipsoid A (being
purely repulsive in character), while for ellipsoid B they are $b_{21}=b_{22}=1.5$, and $b_{23}=1.56$. 
These semiaxes are employed for constructing the shape matrices, which define the repulsive and attractive 
elliptic contact functions.

\subsection{Results}

\begin{figure}[ht]
            \centerline{\includegraphics[width=15cm]{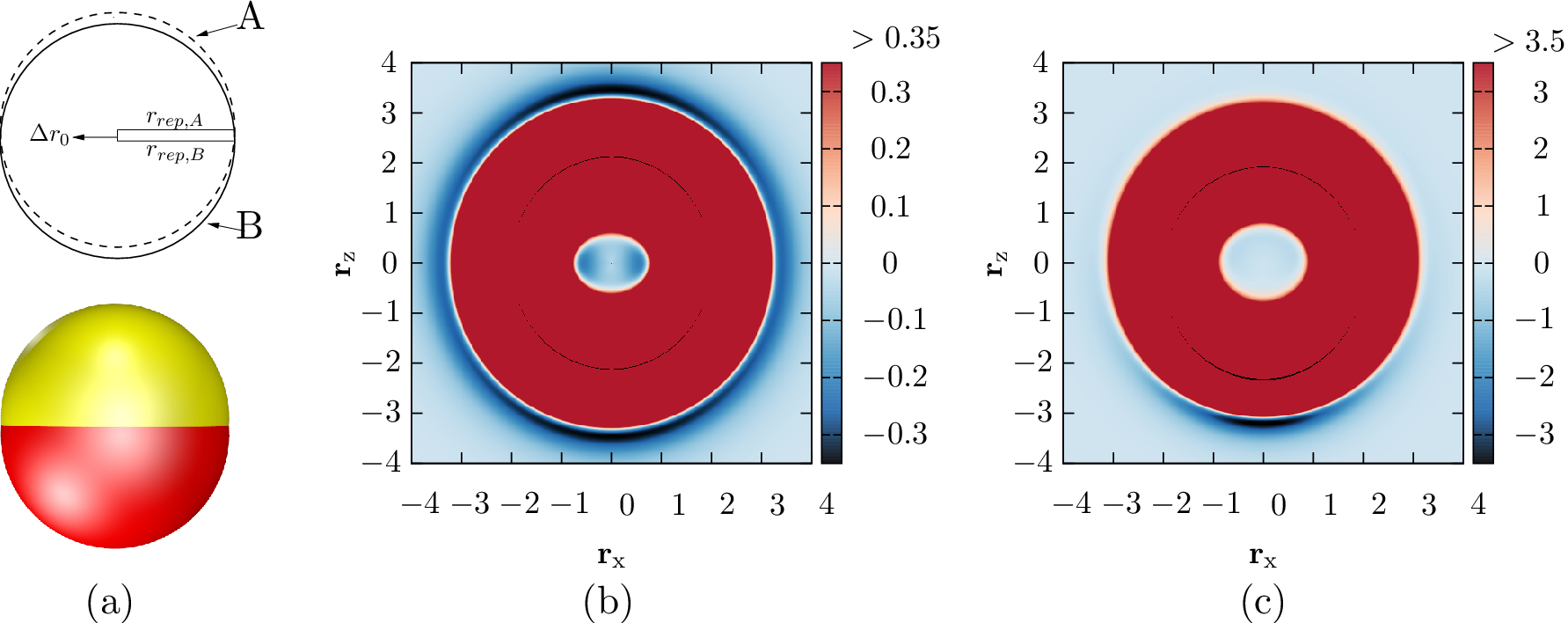}}
	\caption{(a) Schematic view and space-filling representation of the
generic Janus-type building block. Dashed circle and yellow: ellipsoid A (purely repulsive), 
continuous circle and red: ellipsoid B (more attractive along its $z$ axis). $\Delta r_0=0.2$,
$r_{rep,A}=r_{rep,B}=1.5$.
(b) Potential energy surface of two parallel building blocks
confined in the $xz$ plane, with their principal axes aligned with the axes of that plane. 
(c) Same as (b), but with an antiparallel alignment of the 
$z$ axes. Note that the energy range represented in (b) is ten times smaller than that in (c), 
and overlapping configurations with an interaction 
energy outside this range are coloured uniformly with the colour at the top of the range (red).}
\label{fig:surfaces}
\end{figure} 

In this section we introduce a building block composed of two overlapping ellipsoids that strongly favours assembly 
into Bernal spirals (tetrahelices) \cite{boerdijk52,tubularpacking73}. 
Our system behaves similarly to the experimental realisations of Janus particles 
\cite{janus_2011} presented by Chen {\it et al.}, but the underlying energy landscape is likely rather different. In ref.~\cite{janus_2011}
the authors demonstrate that the tetrahelix structures observed at high salt concentrations probably arise due to kinetic 
effects, and other tubular structures such as the 3(0,1,1) helix \cite{tubularpacking73} are not observed because the basic unit of the 
tetrahelix (capped trigonal bipyramid, $N=7$) forms first and it is sufficiently long-lived to aggregate into 
long chains. In contrast, our model ensures that tetrahelix structures are energetically favourable, 
and the interaction profile between building blocks makes the formation of alternative low-energy tubular packings impossible.
Hence we propose a new design, which we predict will guide assembly towards well-defined small 
helical structures very efficiently.

Figure \ref{fig:surfaces}a illustrates our Janus building block. Figures
\ref{fig:surfaces}b and c provide a two-dimensional
representation of the potential energy surfaces resulting from moving two building blocks in the $xz$ plane, 
with their principal axes aligned,
and with the $z$ axes in parallel and antiparallel orientations, respectively. The potential is 
highly attractive at the pole of ellipsoid B furthest away from ellipsoid A. The interaction
range is rather short, and the potential becomes isotropic and decays to zero rapidly as the distance between particles increases.
The displacement by 0.1 distance units of the two ellipsoids is analogous to the experimental method of 
obtaining micrometre-sized Janus particles by coating silica spheres with gold \cite{Chen2011,janus_2011}, since in that case the gold 
coating is thickest at the pole, and gradually decreases towards the equator. The 
hydrophobic interaction itself is determined by the monomolecular layer of alkanethiol applied on the gold coating, and is therefore 
constant on the surface of the patch. However, the net van der Waals interaction experienced by the particles is strongest around 
the pole due to the greater thickness of the gold coating. The main difference between our model and the experimental setup is that 
the deviation from the spherical shape at the poles is about 1\% for the experimental system, while in our case it is about 7\%.
Our potential is also softer than the usual hard sphere-square well representations for experimental colloidal Janus particles. 
Since we did not modify the original PY potential to incorporate Coulombic repulsion, 
our repulsive ellipsoid A has a short range, namely $r^{-12}$.
We find that this repulsion is sufficient to disfavour close contacts between two repulsive ellipsoids, and gives rise to a force that
tends to align two building blocks in an antiparallel fashion. The potential is continuous for every non-overlapping configuration. An 
additional benefit of using the same potential to describe both \lq hemispheres' of the building blocks is that no explicit 
smoothing is required, which would otherwise be necessary to make the interaction profile and its derivative continuous 
around the hydrophobic-hydrophilic interface.
There are discontinuities and unphysical minima in the potential corresponding to highly overlapping configurations, 
but moves that permit such overlaps are 
diagnosed and discarded in our global optimisation and pathway search algorithms. Overlaps between ellipsoidal particles are 
easily detected from
the same elliptic contact function that arises in the energy evaluation.
Such discontinuities and internal wells for overlapping configurations are common for anisotropic potentials \cite{paramonov05,gayberne}.

\begin{figure}[tp]
            \centerline{\includegraphics[width=1.0\textwidth]{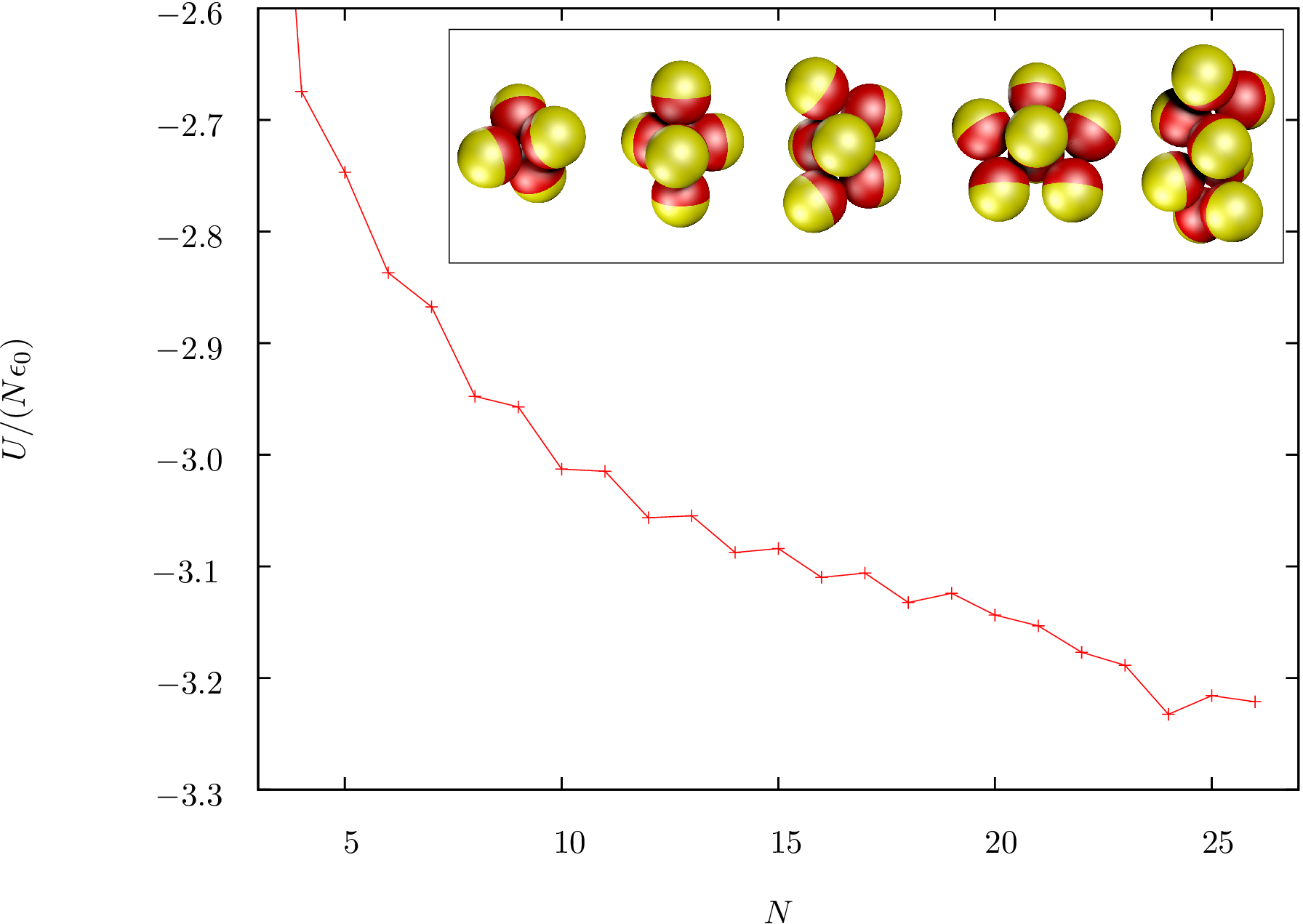}}
\caption{Energy per particle for the predicted global minima in
clusters containing between 4 and 26 Janus particles. The \lq sawtooth' pattern showing a
preference for even number clusters is due to the strongly bound dimer.
Inset: lowest energy structures for small clusters
($N=4$ to 8). 4: tetrahedral, 5: trigonal bipyramidal, 6: capped trigonal bipyramidal (CTBP), 
7: pentagonal bipyramid, 8: Bernal spiral.}
\label{fig:globalminima}
\end{figure}

Our model Janus building blocks strongly prefer to interact {\it via} their attractive poles. 
Since the well depth is not uniform
over the attractive half of the particle, dimerisation is favourable. An additional particle
orients its attractive pole towards the dimer, but the short range and strong directionality of the interaction makes it impossible to 
form a strongly bound cluster. 
The most favourable geometry for the trimer therefore lacks a $C_{3}$ symmetry axis, and the geometry is 
slightly distorted (intercentre distances between the second ellipsoids of each building block are 3.02, 3.08 and 3.08, respectively).
However, when the number of particles in the cluster is even, a complete set of dimers is possible.
This pattern results in hierarchical assembly, where
the dimers themselves behave as larger building blocks,
stacking along their attractive ellipsoids and rotated by 90 degrees. 
When highly symmetric clusters are possible, the strong dimer interactions
can be disrupted if the extra stabilisation from the additional contacts can compete with the dimerisation energy. 
This situation arises for $N=4$ and 5,
with global minima corresponding to tetrahedral and trigonal bipyramidal structures, respectively. 
Increasing the cluster size further destabilises
high symmetry configurations, and assemblies of dimers tend to be preferred. For example,
the global minimum for $N=6$ is a capped trigonal bipyramid (CTBP), not an octahedron, and starting from $N=8$, 
the predicted global minima for every structure with an even number of particles are tetrahelices up to $N=20$.
We emphasise that 
these are not perfect tetrahelix structures, since the tetrahedral units are themselves somewhat distorted, 
the largest difference between the 
edges being around 7\%. The strong preference for dimerisation gives rise to a characteristic 
\lq sawtooth'-pattern in the energy per particle 
{\it versus} cluster size (Figure \ref{fig:globalminima}). The global minima for $N>20$ are ring-like 
structures. There is recent experimental evidence for Janus particles preferring assembly into clusters with even numbers 
in two dimensions \cite{Iwashita2013}, and computational studies on a different Janus building block also show such a 
preference \cite{Preisler2013}, giving rise to similar \lq sawtooth' patterns.

Since there are essentially two types of interactions between the Janus building blocks in a tetrahelix,
namely a stronger and a weaker attraction, assembly of such systems is intrinsically hierarchical through
(i) dimerisation and (ii) association of dimers. Building blocks at 
either end of a finite strand are more weakly bound, so in a bulk system strand growth is preferred if 
there are free dimers available in solution. However, 
increasing the number of particles also allows for certain ring-like structures to arise, built up from tetrahedral units. 
For example, the predicted global
minimum for 24 particles is \lq doughnut'-shaped, corresponding to the first cyclic structure with high symmetry ($D_{6d}$). 
Interestingly, such structures have not yet been observed in experiment \cite{janus_2011}, although 
the main repeating unit in a tetrahelix is the 
CTBP structure \cite{janus_2011}, as in the cyclic global minimum predicted for 24 building blocks.

\begin{figure}[ht]
            \centerline{\includegraphics[width=1.0\textwidth]{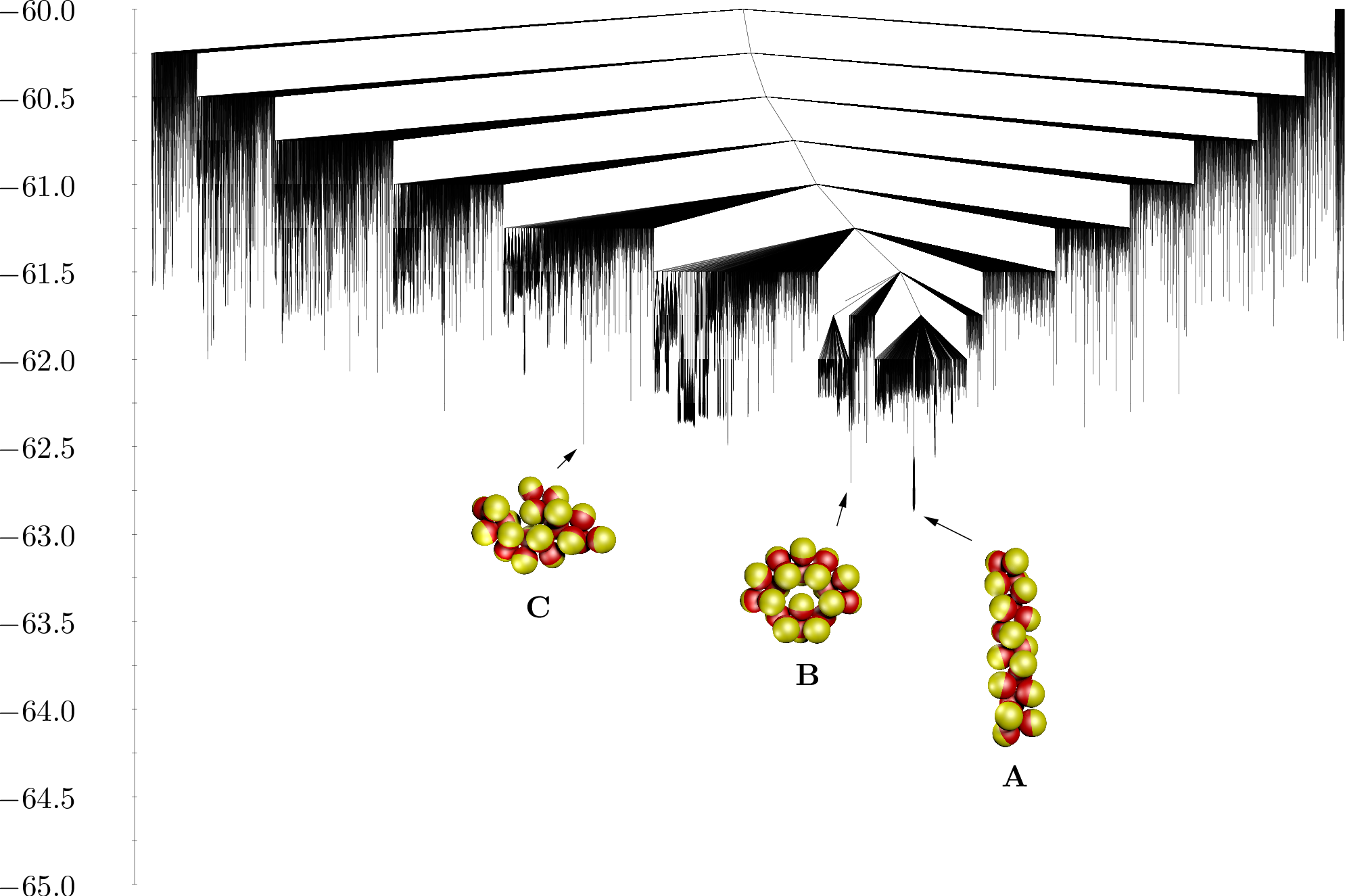}}
        \caption{Disconnectivity graph for $N = 20$ Janus building blocks. The global minimum is a Bernal spiral ({\bf A}), which is very flexible. 
In this graph, more 
than 30 minima around the global minimum
exhibit bent structures,
which interconvert {\it via} \lq hinge' motions. 
The lowest energy kinetic trap ({\bf B}) corresponds 
to a symmetric cyclic structure, while the second-lowest ({\bf C}) is a dimer of a low-energy minimum for $N=10$.}
        \label{fig:disconn20}
\end{figure}

\begin{figure}[ht]
            \centerline{\includegraphics[width=1.0\textwidth]{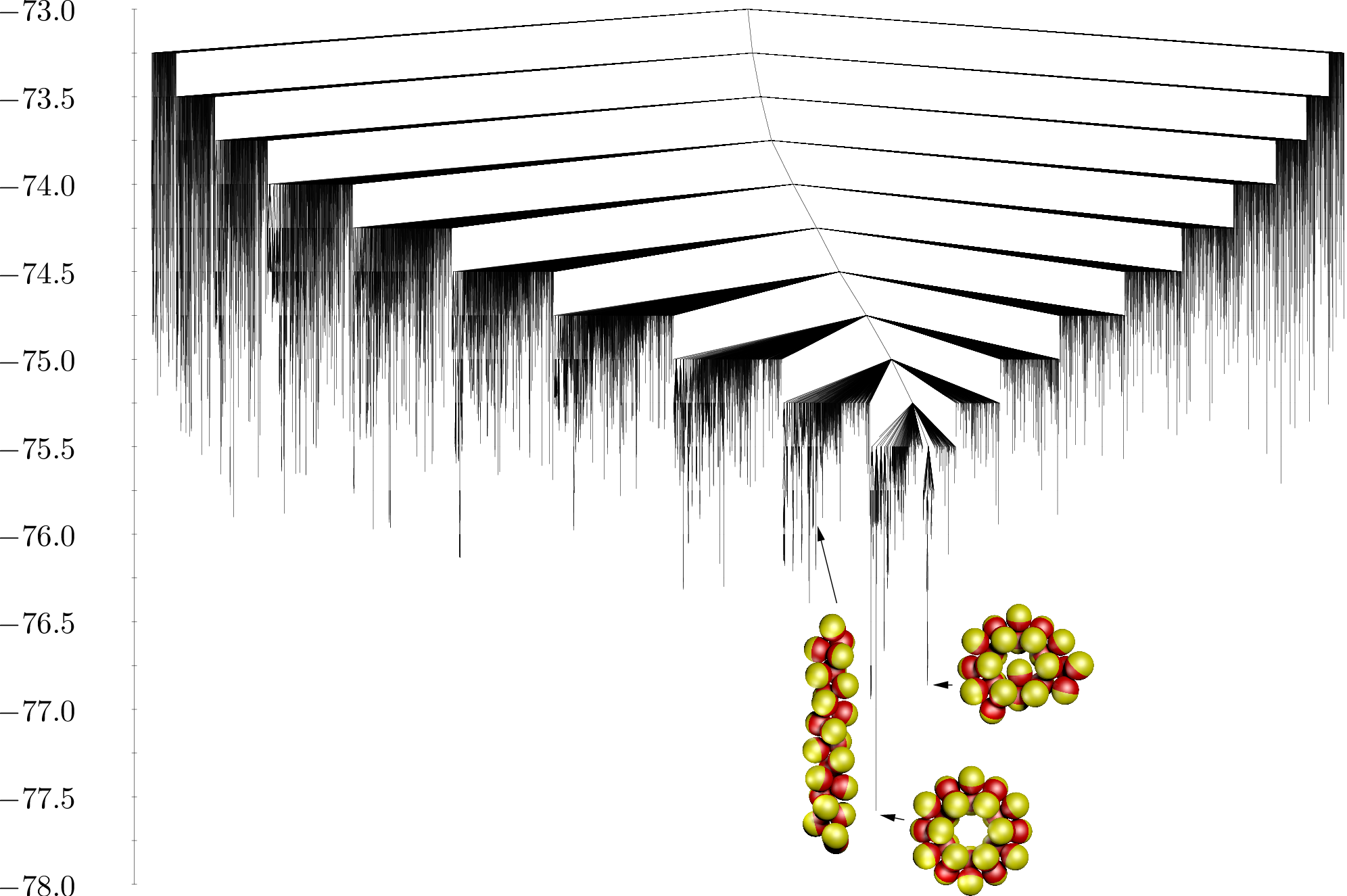}}
        \caption{Disconnectivity graph for $N = 24$ Janus building blocks. 
The global minimum is a ring structure with $D_{6d}$ symmetry.}
        \label{fig:disconn24}
\end{figure}

We have explored the energy landscape more extensively for clusters composed of 20 and 24 particles, 
using discrete path sampling \cite{Wales02,Wales04,Wales06} to grow databases of local minima
and the transition states that connect them. The disconnectivity graphs constructed for the 
two cluster sizes are shown in Figures \ref{fig:disconn20} and \ref{fig:disconn24}, respectively. 
We have used the same energy range, and
positioned the tetrahelical minimum in the same part of the graph, so that the two landscapes can be compared visually.
Enantiomers are lumped together in these graphs.

For $N=20$, we find that the tetrahelical global minimum is very flexible, 
with single transition state rearrangements resulting in bent 
structures that correspond to similar energies. The mechanism is a simple bending motion around a pair of particles 
strongly bound to each other in 
the helical structure, with the dimer acting as a hinge. Only strongly bound dimers act as hinges, 
with their attractive poles almost antiparallel.
Kinetically this is a favourable rearrangement, with relatively low barriers below $0.25\,\epsilon_0$. 
Such \lq hinge'-rearrangements are characteristic low-energy transitions between worm-like structures, 
and are preferred due to the fact that the binding pattern of 
dimers does not change, i.e.~no dimeric binding configuration is disturbed.
The identified \lq hinge'-rearrangements resemble those found for sodium chloride clusters \cite{DoyeW99},
where such rearrangements have relatively high barriers.

\begin{figure}[ht]
            \centerline{\includegraphics[width=1.0\textwidth]{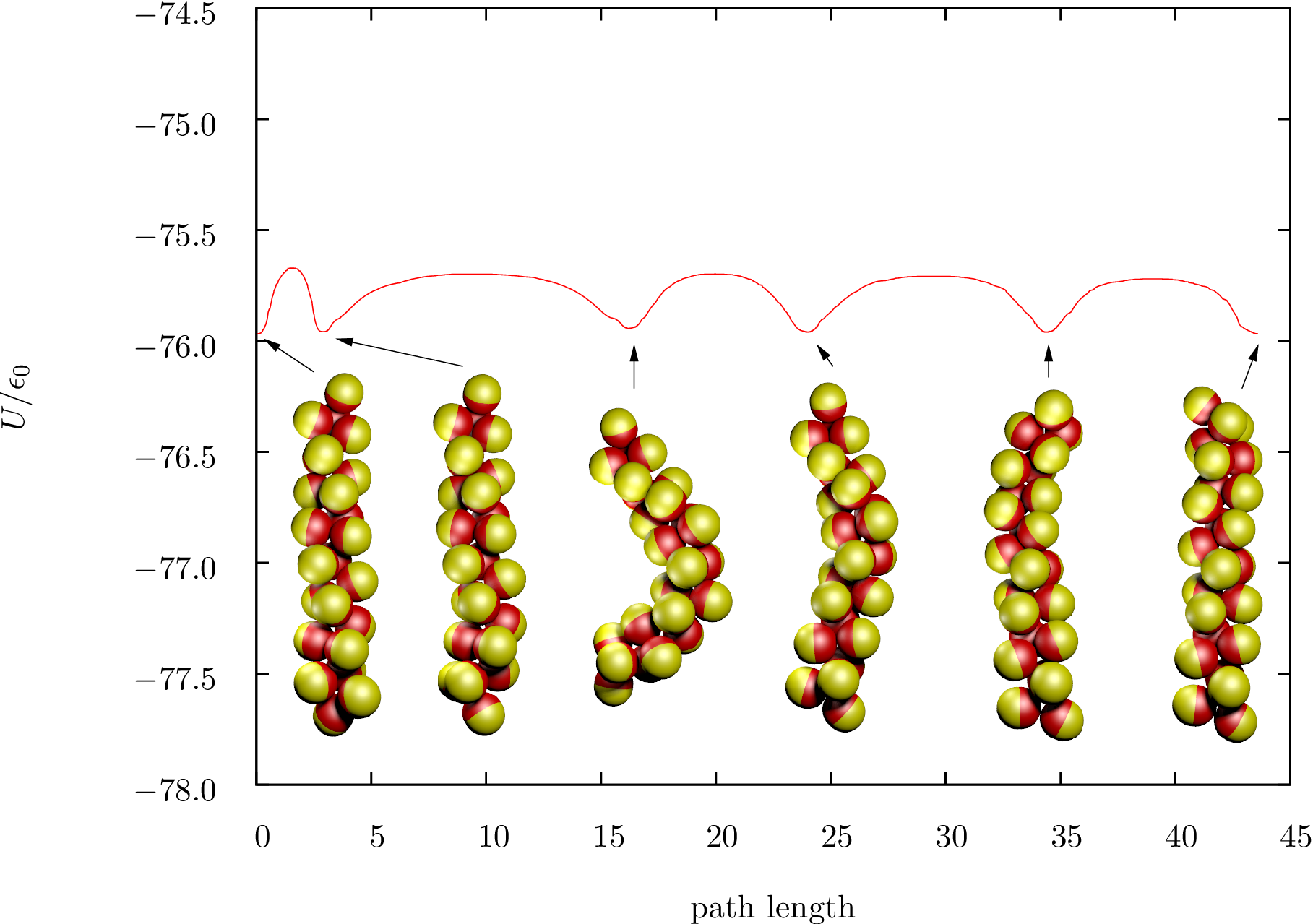}}
        \caption{Predicted fastest pathway for inversion of a 24-particle tetrahelix structure, 
consisting exclusively of \lq hinge' rearrangements. The fourth minimum in the path is in fact 
a dimer composed of a left-handed and a right-handed $N=12$ helix.}
         \label{fig:inversion24}
\end{figure}

\lq Hinge' rearrangement mechanisms are preferred during chirality inversion as well. 
Figure \ref{fig:inversion24} and Supplementary Movie 2 
illustrate one of the fastest pathways between a left-handed and a right-handed $N=24$ helix. All such pathways involve exclusively 
\lq hinge'-motions and a minimum of five transition
states. Such sequential rearrangements have been observed experimentally for smaller clusters \cite{janus_2011}, 
and we see exactly the same type 
of cooperative pathways, resulting in propagation of the change in handedness along the chain.

The disconnectivity graph for $N=20$ contains many kinetic traps, the lowest of which is a ring structure with a symmetry plane ({\bf B}). 
Other low-energy minima that appear as traps are again aggregates of dimers, 
and contain two or more CTBP units. The structure {\bf C} depicted in Figure \ref{fig:disconn20} 
is in fact a dimer of the second-lowest potential energy minimum predicted for the $N=10$ cluster.

\begin{figure}[ht]
            \centerline{\includegraphics[width=1.0\textwidth]{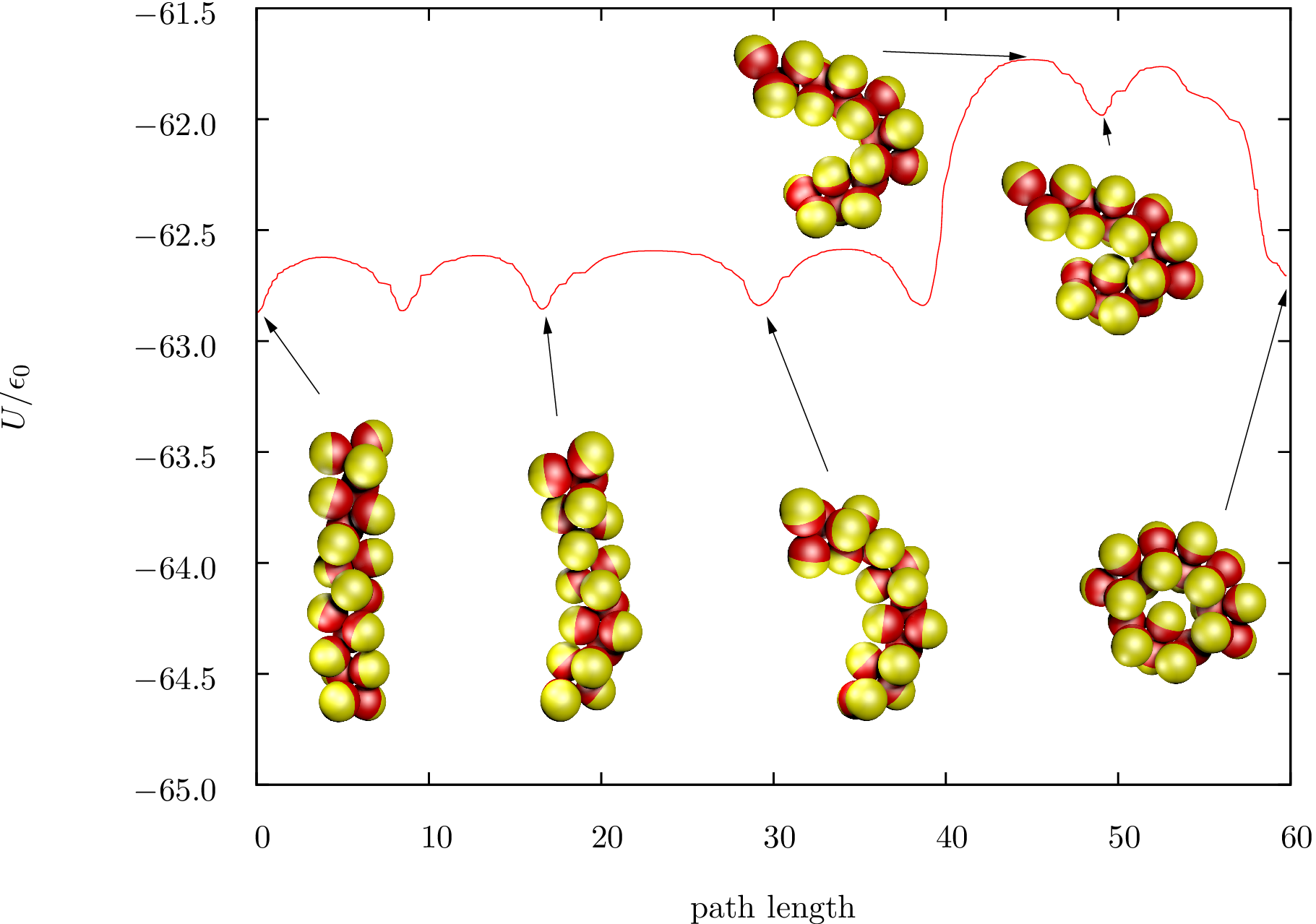}}
        \caption{Fastest pathway between the tetrahelix global minimum and a low-energy closed ring structure. 
The rearrangements with energy barriers of about $0.25\,\epsilon_0$ are all \lq hinge'-motions. 
Structures for selected minima and the highest 
energy transition state are also shown.}
        \label{fig:pathway20}
\end{figure}
The fastest pathway between the helical global minimum and the ring structure 
predicted to act as a kinetic trap
also proceeds through low-energy \lq hinge' mechanisms
(Figure \ref{fig:pathway20}, see also Supplementary Movie 1) up to a point, and the high energy of 
the fifth transition state and the fifth  
minimum along the pathway is due to the torsional strain introduced when the attractive interactions 
between two adjacent dimer units are lost 
(rotation of a dimer around the intercentre vector defined by two loosely bound neighbour particles). 
The final rearrangement in the overall pathway is again a \lq hinge' mechanism.

\begin{figure}[ht]
            \centerline{\includegraphics[width=1.0\textwidth]{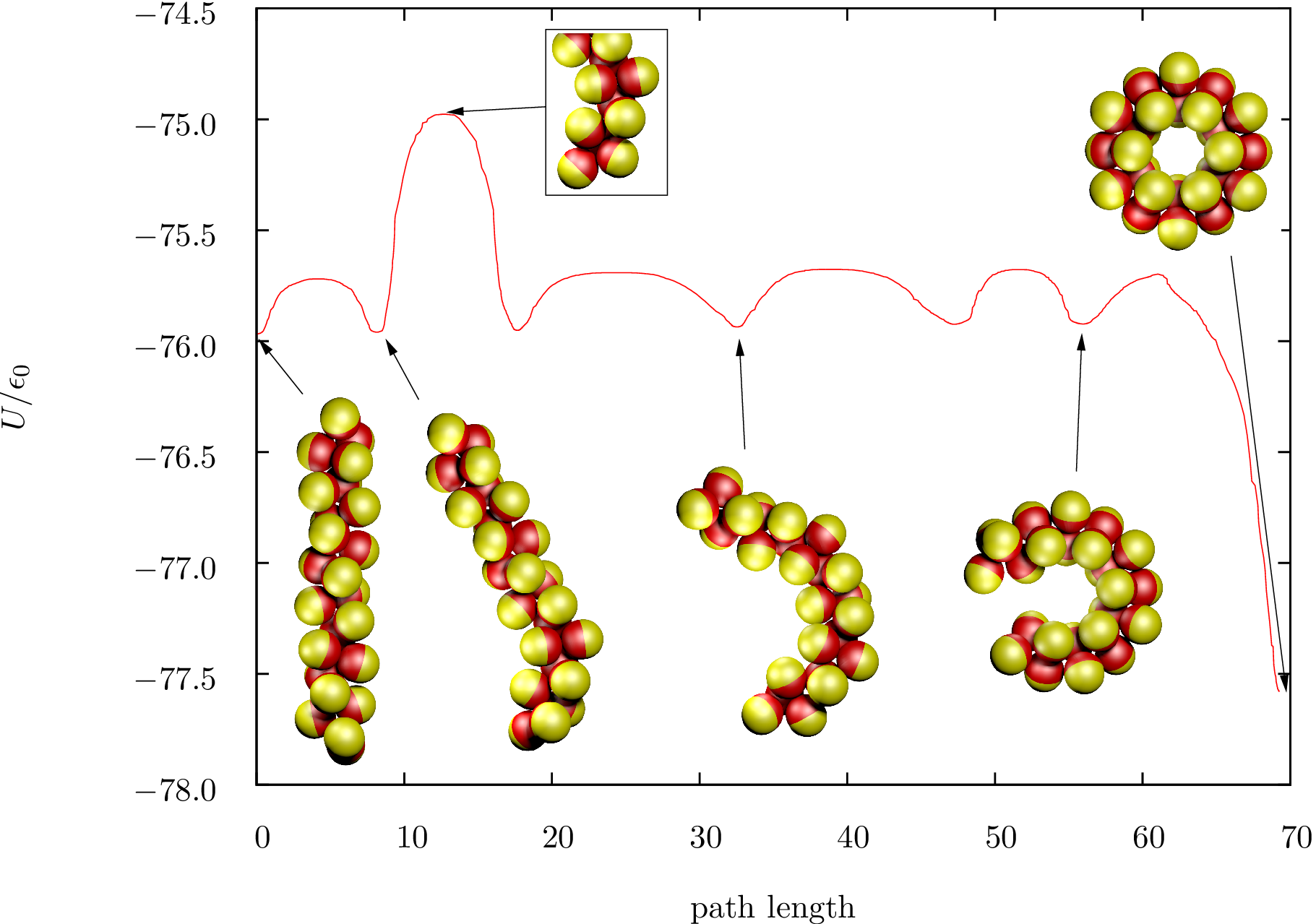}}
        \caption{Fastest pathway between the \lq doughnut'-shaped global minimum of $D_{6d}$ symmetry and the tetrahelix structure. 
The rearrangements
involved with energy barriers of about $0.25\,\epsilon_0$ all correspond to \lq hinge' mechanisms.  
Structures are shown for selected minima. The 
inset shows a part of the cluster in the high energy transition state involving change in the dimerisation pattern.}
        \label{fig:pathway24}
\end{figure}
Helical structures are much higher in energy for $N=24$ clusters 
than the \lq doughnut'-shaped global minimum. Although they exhibit 
the same flexibility as for $N=20$, the energetic separation from the rest of the landscape is not as pronounced, 
and these structures 
can easily undergo ring closures. However, the highly symmetric global minimum cannot be reached through 
simple low-energy \lq hinge' rearrangements. At 
least one step involving a change in dimerisation pattern has to occur at one end of the helix, during which the particles in 
two adjacent dimers rotate cooperatively to change the strong bonding into a weak interaction and {\it vice versa}. 
Such particle rotations result in 
an energy barrier of about 1\,$\epsilon_0$ at the end of the chain, which would be much higher if they 
occurred within the chain. The fastest overall pathway between the helical structure and the global minimum 
is shown in Figure \ref{fig:pathway24} and Supplementary Movie 3.

It would be interesting to investigate how the helical clusters of Janus building blocks designed in ref.~\cite{janus_2011} behave
in dilute solutions, where chain growth is less likely. 
Although chirality inversion has been observed for such helices, all other reported transformations involve
chain growth. The \lq hinge' mechanisms observed experimentally hint at other possible rearrangements for longer helices. In our
model system, the lowest energy transformations involve only \lq hinge' motions. By increasing the temperature, 
barriers could be overcome for the less favoured rearrangements (similar to those described above), 
and in dilute solutions highly symmetric rings might form.

It remains to be seen how our model building blocks behave in a bulk phase. Based on the global optimisation data 
and on the energy landscape analysis for $N=20$ and $24$, a kinetically controlled seeded growth of helices is likely
above 20 particles, by sequential addition of dimers or CTBP units. Assembly of short helices containing even numbers of particles 
will be thermodynamically preferred.

\section{Conclusions}

We have presented two contrasting
theoretical designs for self-assembling helical nanostructures. First 
we described the assembly of charged asymmetric dipolar dumbbells into a helix subject to an 
external magnetic field. Here, helix formation is due to the competition between screened 
electrostatic repulsion and magnetic dipolar interactions. We demonstrate that significant control 
over the helical pitch length (around $30\%$) can be achieved by tuning the balance between 
these two interactions. This tuning is achieved by varying the range of the screened 
electrostatic interactions, which can be realised experimentally by modulating the salt 
concentration of the medium.

We then analysed a model Janus building block that dimerises, where the dimers self-assemble into Bernal spirals, simply by allowing for 
a strong short-range interaction close to the attractive pole of the particle. Large-scale rearrangement mechanisms of such spirals involve sequential \lq hinge' motions.

In agreement with previous work, we find that the formation of complex mesoscopic structures is primarily driven by the anisotropy of building block shape and 
interactions \cite{FejerCW10,fejer_self-assembly_2011}. Our model of rigidly linked ellipsoids can capture a wide range of anisotropy, and the 
self-assembling behaviour for certain sizes is a direct consequence of the building block properties. The model can easily be parameterised to allow 
for larger overlap between building blocks, in order to model \lq softer' Janus particles.

The two models presented here provide two very different approaches to self-assembly 
of helical structures from anisotropic building blocks on the colloidal length scale. In 
both models, helicity is the direct consequence of the shape and interaction anisotropies of 
the building blocks. The overall interaction strength between two particles depends on their 
relative orientation, and on the balance between the repulsive and attractive forces, which 
can be tuned experimentally, for example by varying the ionic strength of the solution. The 
dumbbell model illustrates that varying such experimental conditions makes it possible to 
directly control the helical pitch, while the Janus model provides an example of a 
building-block design that facilitates hierarchical self-assembly into helical structures. 
Both models are relatively simple, and we believe that they are
should be realisable experimentally.

\section{Acknowledgements} 
SNF is a John von Neumann Fellow supported by the European Union and 
the State of Hungary, co-financed by the European Social Fund in the framework of 
T\'AMOP-4.2.4.A/ 2-11/1-2012-0001 \lq National Excellence Program'.
DJW gratefully acknowledges support from the EPSRC and the ERC.


\begin{thebibliography}{63}
\expandafter\ifx\csname natexlab\endcsname\relax\def\natexlab#1{#1}\fi
\expandafter\ifx\csname bibnamefont\endcsname\relax
  \def\bibnamefont#1{#1}\fi
\expandafter\ifx\csname bibfnamefont\endcsname\relax
  \def\bibfnamefont#1{#1}\fi
\expandafter\ifx\csname citenamefont\endcsname\relax
  \def\citenamefont#1{#1}\fi
\expandafter\ifx\csname url\endcsname\relax
  \def\url#1{\texttt{#1}}\fi
\expandafter\ifx\csname urlprefix\endcsname\relax\def\urlprefix{URL }\fi
\providecommand{\bibinfo}[2]{#2}
\providecommand{\eprint}[2][]{\url{#2}}

\bibitem[{\citenamefont{Percec et~al.}(2002)\citenamefont{Percec, Glodde, Bera,
  Miura, I, Singer, Balagurusamy, Heiney, Schnell, A et~al.}}]{PercecETAL02}
\bibinfo{author}{\bibfnamefont{V.}~\bibnamefont{Percec}},
  \bibinfo{author}{\bibfnamefont{M.}~\bibnamefont{Glodde}},
  \bibinfo{author}{\bibfnamefont{T.~K.} \bibnamefont{Bera}},
  \bibinfo{author}{\bibfnamefont{Y.}~\bibnamefont{Miura}},
  \bibinfo{author}{\bibfnamefont{I.~S.} \bibnamefont{I}},
  \bibinfo{author}{\bibfnamefont{K.~D.} \bibnamefont{Singer}},
  \bibinfo{author}{\bibfnamefont{V.~S.} \bibnamefont{Balagurusamy}},
  \bibinfo{author}{\bibfnamefont{P.~A.} \bibnamefont{Heiney}},
  \bibinfo{author}{\bibfnamefont{I.}~\bibnamefont{Schnell}},
  \bibinfo{author}{\bibfnamefont{A.~R.} \bibnamefont{A}}, \bibnamefont{et~al.},
  \bibinfo{journal}{Nature} \textbf{\bibinfo{volume}{419}},
  \bibinfo{pages}{384} (\bibinfo{year}{2002}).

\bibitem[{\citenamefont{Yashima et~al.}(2009)\citenamefont{Yashima, Maeda,
  Iida, Furusho, and Nagai}}]{YashimaMIFN09}
\bibinfo{author}{\bibfnamefont{E.}~\bibnamefont{Yashima}},
  \bibinfo{author}{\bibfnamefont{K.}~\bibnamefont{Maeda}},
  \bibinfo{author}{\bibfnamefont{H.}~\bibnamefont{Iida}},
  \bibinfo{author}{\bibfnamefont{Y.}~\bibnamefont{Furusho}}, \bibnamefont{and}
  \bibinfo{author}{\bibfnamefont{K.}~\bibnamefont{Nagai}},
  \bibinfo{journal}{Chem. Rev.} \textbf{\bibinfo{volume}{109}},
  \bibinfo{pages}{6102} (\bibinfo{year}{2009}).

\bibitem[{\citenamefont{Pijper and Feringa}(2008)}]{PijperF08}
\bibinfo{author}{\bibfnamefont{D.}~\bibnamefont{Pijper}} \bibnamefont{and}
  \bibinfo{author}{\bibfnamefont{B.~L.} \bibnamefont{Feringa}},
  \bibinfo{journal}{Soft Matter} \textbf{\bibinfo{volume}{4}},
  \bibinfo{pages}{1349} (\bibinfo{year}{2008}).

\bibitem[{\citenamefont{Gier et~al.}(1998)\citenamefont{Gier, Bu, Feng, and
  Stucky}}]{GierBFS98}
\bibinfo{author}{\bibfnamefont{T.~E.} \bibnamefont{Gier}},
  \bibinfo{author}{\bibfnamefont{X.}~\bibnamefont{Bu}},
  \bibinfo{author}{\bibfnamefont{P.}~\bibnamefont{Feng}}, \bibnamefont{and}
  \bibinfo{author}{\bibfnamefont{G.~D.} \bibnamefont{Stucky}},
  \bibinfo{journal}{Nature} \textbf{\bibinfo{volume}{395}},
  \bibinfo{pages}{154} (\bibinfo{year}{1998}).

\bibitem[{\citenamefont{Lehn}(2002)}]{Lehn02}
\bibinfo{author}{\bibfnamefont{J.-M.} \bibnamefont{Lehn}},
  \bibinfo{journal}{Science} \textbf{\bibinfo{volume}{295}},
  \bibinfo{pages}{2400} (\bibinfo{year}{2002}).

\bibitem[{\citenamefont{Glotzer and Solomon}(2007)}]{GlotzerS07}
\bibinfo{author}{\bibfnamefont{S.~C.} \bibnamefont{Glotzer}} \bibnamefont{and}
  \bibinfo{author}{\bibfnamefont{M.~J.} \bibnamefont{Solomon}},
  \bibinfo{journal}{Nature Mater.} \textbf{\bibinfo{volume}{6}},
  \bibinfo{pages}{557} (\bibinfo{year}{2007}).

\bibitem[{\citenamefont{Sacanna et~al.}(2010)\citenamefont{Sacanna, Irvine,
  Chaikin, and Pine}}]{SacannaICP10}
\bibinfo{author}{\bibfnamefont{S.}~\bibnamefont{Sacanna}},
  \bibinfo{author}{\bibfnamefont{W.~T.~M.} \bibnamefont{Irvine}},
  \bibinfo{author}{\bibfnamefont{P.~M.} \bibnamefont{Chaikin}},
  \bibnamefont{and} \bibinfo{author}{\bibfnamefont{D.~J.} \bibnamefont{Pine}},
  \bibinfo{journal}{Nature} \textbf{\bibinfo{volume}{464}},
  \bibinfo{pages}{575} (\bibinfo{year}{2010}).

\bibitem[{\citenamefont{Min et~al.}(2008)\citenamefont{Min, Akbulut,
  Kristiansen, Golan, and Israelachvili}}]{MinAKGI08}
\bibinfo{author}{\bibfnamefont{Y.}~\bibnamefont{Min}},
  \bibinfo{author}{\bibfnamefont{M.}~\bibnamefont{Akbulut}},
  \bibinfo{author}{\bibfnamefont{K.}~\bibnamefont{Kristiansen}},
  \bibinfo{author}{\bibfnamefont{Y.}~\bibnamefont{Golan}}, \bibnamefont{and}
  \bibinfo{author}{\bibfnamefont{J.}~\bibnamefont{Israelachvili}},
  \bibinfo{journal}{Nature Mater.} \textbf{\bibinfo{volume}{7}},
  \bibinfo{pages}{527} (\bibinfo{year}{2008}).

\bibitem[{\citenamefont{Chakrabarti et~al.}(2009)\citenamefont{Chakrabarti,
  Fejer, and Wales}}]{ChakrabartiFW09}
\bibinfo{author}{\bibfnamefont{D.}~\bibnamefont{Chakrabarti}},
  \bibinfo{author}{\bibfnamefont{S.~N.} \bibnamefont{Fejer}}, \bibnamefont{and}
  \bibinfo{author}{\bibfnamefont{D.~J.} \bibnamefont{Wales}},
  \bibinfo{journal}{Proc. Natl. Acad. Sci. USA} \textbf{\bibinfo{volume}{106}},
  \bibinfo{pages}{20164} (\bibinfo{year}{2009}).

\bibitem[{\citenamefont{Zerrouki et~al.}(2008)\citenamefont{Zerrouki, Boudri,
  Pine, Chaikin, and Bibette}}]{ZerroukiBPCB08}
\bibinfo{author}{\bibfnamefont{D.}~\bibnamefont{Zerrouki}},
  \bibinfo{author}{\bibfnamefont{J.}~\bibnamefont{Boudri}},
  \bibinfo{author}{\bibfnamefont{D.}~\bibnamefont{Pine}},
  \bibinfo{author}{\bibfnamefont{P.}~\bibnamefont{Chaikin}}, \bibnamefont{and}
  \bibinfo{author}{\bibfnamefont{J.}~\bibnamefont{Bibette}},
  \bibinfo{journal}{Nature} \textbf{\bibinfo{volume}{455}},
  \bibinfo{pages}{380} (\bibinfo{year}{2008}).

\bibitem[{\citenamefont{Watson and Crick}(1953)}]{WatsonC53}
\bibinfo{author}{\bibfnamefont{J.~D.} \bibnamefont{Watson}} \bibnamefont{and}
  \bibinfo{author}{\bibfnamefont{F.~H.~C.} \bibnamefont{Crick}},
  \bibinfo{journal}{Nature} \textbf{\bibinfo{volume}{171}},
  \bibinfo{pages}{737} (\bibinfo{year}{1953}).

\bibitem[{\citenamefont{Howson et~al.}(2012)\citenamefont{Howson, Bolhuis,
  Brabec, Clarkson, Malina, Rodger, and Scott}}]{HowsonETAL12}
\bibinfo{author}{\bibfnamefont{S.~E.} \bibnamefont{Howson}},
  \bibinfo{author}{\bibfnamefont{A.}~\bibnamefont{Bolhuis}},
  \bibinfo{author}{\bibfnamefont{V.}~\bibnamefont{Brabec}},
  \bibinfo{author}{\bibfnamefont{G.~J.} \bibnamefont{Clarkson}},
  \bibinfo{author}{\bibfnamefont{J.}~\bibnamefont{Malina}},
  \bibinfo{author}{\bibfnamefont{A.}~\bibnamefont{Rodger}}, \bibnamefont{and}
  \bibinfo{author}{\bibfnamefont{P.}~\bibnamefont{Scott}},
  \bibinfo{journal}{Nature Chem.} \textbf{\bibinfo{volume}{4}},
  \bibinfo{pages}{31} (\bibinfo{year}{2012}).

\bibitem[{\citenamefont{Srivastava et~al.}(2010)\citenamefont{Srivastava,
  Santos, Critchley, Kim, Podsiadlo, Sun, Lee, Xu, Lilly, Glotzer
  et~al.}}]{SrivastavaETAL10}
\bibinfo{author}{\bibfnamefont{S.}~\bibnamefont{Srivastava}},
  \bibinfo{author}{\bibfnamefont{A.}~\bibnamefont{Santos}},
  \bibinfo{author}{\bibfnamefont{K.}~\bibnamefont{Critchley}},
  \bibinfo{author}{\bibfnamefont{K.-S.} \bibnamefont{Kim}},
  \bibinfo{author}{\bibfnamefont{P.}~\bibnamefont{Podsiadlo}},
  \bibinfo{author}{\bibfnamefont{K.}~\bibnamefont{Sun}},
  \bibinfo{author}{\bibfnamefont{J.}~\bibnamefont{Lee}},
  \bibinfo{author}{\bibfnamefont{C.}~\bibnamefont{Xu}},
  \bibinfo{author}{\bibfnamefont{D.}~\bibnamefont{Lilly}},
  \bibinfo{author}{\bibfnamefont{S.~C.} \bibnamefont{Glotzer}},
  \bibnamefont{et~al.}, \bibinfo{journal}{Science}
  \textbf{\bibinfo{volume}{327}}, \bibinfo{pages}{1355} (\bibinfo{year}{2010}).

\bibitem[{\citenamefont{Hong et~al.}(2008)\citenamefont{Hong, Cacciuto,
  Luijten, and Granick}}]{janus_langmuir_2008}
\bibinfo{author}{\bibfnamefont{L.}~\bibnamefont{Hong}},
  \bibinfo{author}{\bibfnamefont{A.}~\bibnamefont{Cacciuto}},
  \bibinfo{author}{\bibfnamefont{E.}~\bibnamefont{Luijten}}, \bibnamefont{and}
  \bibinfo{author}{\bibfnamefont{S.}~\bibnamefont{Granick}},
  \bibinfo{journal}{Langmuir} \textbf{\bibinfo{volume}{24}},
  \bibinfo{pages}{621} (\bibinfo{year}{2008}).

\bibitem[{\citenamefont{Li et~al.}(2012)\citenamefont{Li, Lu, Sun, and
  An}}]{janus_softmatter_2012}
\bibinfo{author}{\bibfnamefont{Z.-W.} \bibnamefont{Li}},
  \bibinfo{author}{\bibfnamefont{Z.-Y.} \bibnamefont{Lu}},
  \bibinfo{author}{\bibfnamefont{Z.-Y.} \bibnamefont{Sun}}, \bibnamefont{and}
  \bibinfo{author}{\bibfnamefont{L.-J.} \bibnamefont{An}},
  \bibinfo{journal}{Soft Matter} \textbf{\bibinfo{volume}{8}},
  \bibinfo{pages}{6693} (\bibinfo{year}{2012}).

\bibitem[{\citenamefont{Liu et~al.}(2012)\citenamefont{Liu, Li, Perez, Gunton,
  and Brett}}]{janus_langmuir_2012}
\bibinfo{author}{\bibfnamefont{Y.}~\bibnamefont{Liu}},
  \bibinfo{author}{\bibfnamefont{W.}~\bibnamefont{Li}},
  \bibinfo{author}{\bibfnamefont{T.}~\bibnamefont{Perez}},
  \bibinfo{author}{\bibfnamefont{J.~D.} \bibnamefont{Gunton}},
  \bibnamefont{and} \bibinfo{author}{\bibfnamefont{G.}~\bibnamefont{Brett}},
  \bibinfo{journal}{Langmuir} \textbf{\bibinfo{volume}{28}}, \bibinfo{pages}{3}
  (\bibinfo{year}{2012}).

\bibitem[{\citenamefont{Li and Gunton}(2013)}]{janus_langmuir_2013}
\bibinfo{author}{\bibfnamefont{W.}~\bibnamefont{Li}} \bibnamefont{and}
  \bibinfo{author}{\bibfnamefont{J.~D.} \bibnamefont{Gunton}},
  \bibinfo{journal}{Langmuir} \textbf{\bibinfo{volume}{29}},
  \bibinfo{pages}{8517} (\bibinfo{year}{2013}).

\bibitem[{\citenamefont{Chen et~al.}(2011{\natexlab{a}})\citenamefont{Chen,
  Whitmer, Jiang, Bae, Luijten, and Granick}}]{janus_2011}
\bibinfo{author}{\bibfnamefont{Q.}~\bibnamefont{Chen}},
  \bibinfo{author}{\bibfnamefont{J.~K.} \bibnamefont{Whitmer}},
  \bibinfo{author}{\bibfnamefont{S.}~\bibnamefont{Jiang}},
  \bibinfo{author}{\bibfnamefont{S.~C.} \bibnamefont{Bae}},
  \bibinfo{author}{\bibfnamefont{E.}~\bibnamefont{Luijten}}, \bibnamefont{and}
  \bibinfo{author}{\bibfnamefont{S.}~\bibnamefont{Granick}},
  \bibinfo{journal}{Science} \textbf{\bibinfo{volume}{331}},
  \bibinfo{pages}{199} (\bibinfo{year}{2011}{\natexlab{a}}).

\bibitem[{\citenamefont{Campbell et~al.}(2005)\citenamefont{Campbell, Anderson,
  van Duijneveldt, and Bartlett}}]{Campbell2005}
\bibinfo{author}{\bibfnamefont{A.~I.} \bibnamefont{Campbell}},
  \bibinfo{author}{\bibfnamefont{V.~J.} \bibnamefont{Anderson}},
  \bibinfo{author}{\bibfnamefont{J.~S.} \bibnamefont{van Duijneveldt}},
  \bibnamefont{and} \bibinfo{author}{\bibfnamefont{P.}~\bibnamefont{Bartlett}},
  \bibinfo{journal}{Phys. Rev. Lett.} \textbf{\bibinfo{volume}{94}},
  \bibinfo{pages}{208301} (\bibinfo{year}{2005}).

\bibitem[{\citenamefont{Mossa et~al.}(2004)\citenamefont{Mossa, Sciortino,
  Tartaglia, and Zaccarelli}}]{Mossa2004}
\bibinfo{author}{\bibfnamefont{S.}~\bibnamefont{Mossa}},
  \bibinfo{author}{\bibfnamefont{F.}~\bibnamefont{Sciortino}},
  \bibinfo{author}{\bibfnamefont{P.}~\bibnamefont{Tartaglia}},
  \bibnamefont{and}
  \bibinfo{author}{\bibfnamefont{E.}~\bibnamefont{Zaccarelli}},
  \bibinfo{journal}{Langmuir} \textbf{\bibinfo{volume}{20}},
  \bibinfo{pages}{10756} (\bibinfo{year}{2004}).

\bibitem[{\citenamefont{Sciortino et~al.}(2005)\citenamefont{Sciortino,
  Tartaglia, and Zaccarelli}}]{Sciortino2005}
\bibinfo{author}{\bibfnamefont{F.}~\bibnamefont{Sciortino}},
  \bibinfo{author}{\bibfnamefont{P.}~\bibnamefont{Tartaglia}},
  \bibnamefont{and}
  \bibinfo{author}{\bibfnamefont{E.}~\bibnamefont{Zaccarelli}},
  \bibinfo{journal}{J. Phys. Chem. B} \textbf{\bibinfo{volume}{109}},
  \bibinfo{pages}{21942} (\bibinfo{year}{2005}).

\bibitem[{\citenamefont{Li and Scheraga}(1987)}]{Scheraga87}
\bibinfo{author}{\bibfnamefont{Z.}~\bibnamefont{Li}} \bibnamefont{and}
  \bibinfo{author}{\bibfnamefont{H.~A.} \bibnamefont{Scheraga}},
  \bibinfo{journal}{\pnas} \textbf{\bibinfo{volume}{84}}, \bibinfo{pages}{6611}
  (\bibinfo{year}{1987}).

\bibitem[{\citenamefont{Wales and Doye}(1997)}]{WalesD97}
\bibinfo{author}{\bibfnamefont{D.~J.} \bibnamefont{Wales}} \bibnamefont{and}
  \bibinfo{author}{\bibfnamefont{J.~P.~K.} \bibnamefont{Doye}},
  \bibinfo{journal}{\jpca} \textbf{\bibinfo{volume}{101}},
  \bibinfo{pages}{5111} (\bibinfo{year}{1997}).

\bibitem[{\citenamefont{Wales and Scheraga}(1999)}]{waless99}
\bibinfo{author}{\bibfnamefont{D.~J.} \bibnamefont{Wales}} \bibnamefont{and}
  \bibinfo{author}{\bibfnamefont{H.~A.} \bibnamefont{Scheraga}},
  \bibinfo{journal}{Science} \textbf{\bibinfo{volume}{285}},
  \bibinfo{pages}{1368} (\bibinfo{year}{1999}).

\bibitem[{\citenamefont{Wales}({\natexlab{a}})}]{GMIN}
\bibinfo{author}{\bibfnamefont{D.~J.} \bibnamefont{Wales}},
  \emph{\bibinfo{title}{{GMIN}: A program for finding global minima and
  calculating thermodynamic properties from basin-sampling,}},
  \bibinfo{address}{http://www-wales.ch.cam.ac.uk/{GMIN}} (????{\natexlab{a}}).

\bibitem[{\citenamefont{Wales}(2002)}]{Wales02}
\bibinfo{author}{\bibfnamefont{D.~J.} \bibnamefont{Wales}},
  \bibinfo{journal}{Mol. Phys.} \textbf{\bibinfo{volume}{100}},
  \bibinfo{pages}{3285} (\bibinfo{year}{2002}).

\bibitem[{\citenamefont{Wales}(2004)}]{Wales04}
\bibinfo{author}{\bibfnamefont{D.~J.} \bibnamefont{Wales}},
  \bibinfo{journal}{Mol. Phys.} \textbf{\bibinfo{volume}{102}},
  \bibinfo{pages}{891} (\bibinfo{year}{2004}).

\bibitem[{\citenamefont{Wales}(2006)}]{Wales06}
\bibinfo{author}{\bibfnamefont{D.~J.} \bibnamefont{Wales}},
  \bibinfo{journal}{Int. Rev. Phys. Chem.} \textbf{\bibinfo{volume}{25}},
  \bibinfo{pages}{237} (\bibinfo{year}{2006}).

\bibitem[{\citenamefont{Wales}({\natexlab{b}})}]{optim}
\bibinfo{author}{\bibfnamefont{D.~J.} \bibnamefont{Wales}},
  \emph{\bibinfo{title}{{OPTIM}: A program for characterising stationary points
  and reaction pathways, {\rm http://www-wales.ch.cam.ac.uk/}{{\rm OPTIM}}}},
  \bibinfo{address}{http://www-wales.ch.cam.ac.uk/OPTIM} (????{\natexlab{b}}).

\bibitem[{\citenamefont{Wales}({\natexlab{c}})}]{pathsample}
\bibinfo{author}{\bibfnamefont{D.~J.} \bibnamefont{Wales}},
  \emph{\bibinfo{title}{{PATHSAMPLE}: A program for refining and analysing
  kinetic transition networks}},
  \bibinfo{address}{http://www-wales.ch.cam.ac.uk/{PATHSAMPLE}}
  (????{\natexlab{c}}).

\bibitem[{\citenamefont{Trygubenko and
  Wales}(2004{\natexlab{a}})}]{trygubenkow04}
\bibinfo{author}{\bibfnamefont{S.~A.} \bibnamefont{Trygubenko}}
  \bibnamefont{and} \bibinfo{author}{\bibfnamefont{D.~J.} \bibnamefont{Wales}},
  \bibinfo{journal}{J. Chem. Phys.} \textbf{\bibinfo{volume}{120}},
  \bibinfo{pages}{2082} (\bibinfo{year}{2004}{\natexlab{a}}).

\bibitem[{\citenamefont{Henkelman and J\'onsson}(1999)}]{henkelmanj99}
\bibinfo{author}{\bibfnamefont{G.}~\bibnamefont{Henkelman}} \bibnamefont{and}
  \bibinfo{author}{\bibfnamefont{H.}~\bibnamefont{J\'onsson}},
  \bibinfo{journal}{\jcp} \textbf{\bibinfo{volume}{111}}, \bibinfo{pages}{7010}
  (\bibinfo{year}{1999}).

\bibitem[{\citenamefont{Henkelman et~al.}(2000)\citenamefont{Henkelman,
  Uberuaga, and J\'onsson}}]{HenkelmanUJ00}
\bibinfo{author}{\bibfnamefont{G.}~\bibnamefont{Henkelman}},
  \bibinfo{author}{\bibfnamefont{B.~P.} \bibnamefont{Uberuaga}},
  \bibnamefont{and}
  \bibinfo{author}{\bibfnamefont{H.}~\bibnamefont{J\'onsson}},
  \bibinfo{journal}{J. Chem. Phys.} \textbf{\bibinfo{volume}{113}},
  \bibinfo{pages}{9901} (\bibinfo{year}{2000}).

\bibitem[{\citenamefont{Henkelman and J\'onsson}(2000)}]{HenkelmanJ00}
\bibinfo{author}{\bibfnamefont{G.}~\bibnamefont{Henkelman}} \bibnamefont{and}
  \bibinfo{author}{\bibfnamefont{H.}~\bibnamefont{J\'onsson}},
  \bibinfo{journal}{J. Chem. Phys.} \textbf{\bibinfo{volume}{113}},
  \bibinfo{pages}{9978} (\bibinfo{year}{2000}).

\bibitem[{\citenamefont{Trygubenko and
  Wales}(2004{\natexlab{b}})}]{trygubenko04}
\bibinfo{author}{\bibfnamefont{S.~A.} \bibnamefont{Trygubenko}}
  \bibnamefont{and} \bibinfo{author}{\bibfnamefont{D.~J.} \bibnamefont{Wales}},
  \bibinfo{journal}{J. Chem. Phys.} \textbf{\bibinfo{volume}{120}},
  \bibinfo{pages}{2082} (\bibinfo{year}{2004}{\natexlab{b}}).

\bibitem[{\citenamefont{Munro and Wales}(1999)}]{MunroW99}
\bibinfo{author}{\bibfnamefont{L.~J.} \bibnamefont{Munro}} \bibnamefont{and}
  \bibinfo{author}{\bibfnamefont{D.~J.} \bibnamefont{Wales}},
  \bibinfo{journal}{Phys. Rev. B} \textbf{\bibinfo{volume}{59}},
  \bibinfo{pages}{3969} (\bibinfo{year}{1999}).

\bibitem[{\citenamefont{Dijkstra}(1959)}]{Dijkstra59}
\bibinfo{author}{\bibfnamefont{E.}~\bibnamefont{Dijkstra}},
  \bibinfo{journal}{Numer. Math.} \textbf{\bibinfo{volume}{1}},
  \bibinfo{pages}{269} (\bibinfo{year}{1959}).

\bibitem[{\citenamefont{Becker and Karplus}(1997)}]{BeckerK97}
\bibinfo{author}{\bibfnamefont{O.~M.} \bibnamefont{Becker}} \bibnamefont{and}
  \bibinfo{author}{\bibfnamefont{M.}~\bibnamefont{Karplus}},
  \bibinfo{journal}{J. Chem. Phys.} \textbf{\bibinfo{volume}{106}},
  \bibinfo{pages}{1495} (\bibinfo{year}{1997}).

\bibitem[{\citenamefont{Wales et~al.}(1998)\citenamefont{Wales, Miller, and
  Walsh}}]{WalesMW98}
\bibinfo{author}{\bibfnamefont{D.~J.} \bibnamefont{Wales}},
  \bibinfo{author}{\bibfnamefont{M.~A.} \bibnamefont{Miller}},
  \bibnamefont{and} \bibinfo{author}{\bibfnamefont{T.~R.} \bibnamefont{Walsh}},
  \bibinfo{journal}{Nature} \textbf{\bibinfo{volume}{394}},
  \bibinfo{pages}{758} (\bibinfo{year}{1998}).

\bibitem[{\citenamefont{Wales}(2003)}]{Wales03}
\bibinfo{author}{\bibfnamefont{D.~J.} \bibnamefont{Wales}},
  \emph{\bibinfo{title}{{Energy Landscapes: Applications to Clusters,
  Biomolecules and Glasses}}} (\bibinfo{publisher}{Cambridge University Press},
  \bibinfo{year}{2003}).

\bibitem[{\citenamefont{Wales}(2010)}]{Wales10a}
\bibinfo{author}{\bibfnamefont{D.~J.} \bibnamefont{Wales}},
  \bibinfo{journal}{Curr.~Op.~Struct.~Biol.} \textbf{\bibinfo{volume}{20}},
  \bibinfo{pages}{3} (\bibinfo{year}{2010}).

\bibitem[{\citenamefont{Wales}(2012)}]{Wales12}
\bibinfo{author}{\bibfnamefont{D.~J.} \bibnamefont{Wales}},
  \bibinfo{journal}{Phil. Trans. Roy. Soc. A} \textbf{\bibinfo{volume}{370}},
  \bibinfo{pages}{2877} (\bibinfo{year}{2012}).

\bibitem[{\citenamefont{Robbins et~al.}(1988)\citenamefont{Robbins, Kremer, and
  Grest}}]{RobbinsKG92}
\bibinfo{author}{\bibfnamefont{M.~O.} \bibnamefont{Robbins}},
  \bibinfo{author}{\bibfnamefont{K.}~\bibnamefont{Kremer}}, \bibnamefont{and}
  \bibinfo{author}{\bibfnamefont{G.~S.} \bibnamefont{Grest}},
  \bibinfo{journal}{J. Chem. Phys.} \textbf{\bibinfo{volume}{88}},
  \bibinfo{pages}{3286} (\bibinfo{year}{1988}).

\bibitem[{\citenamefont{{L\"{o}wen} and Kramposthuber}(1993)}]{LowenK93}
\bibinfo{author}{\bibfnamefont{H.}~\bibnamefont{{L\"{o}wen}}} \bibnamefont{and}
  \bibinfo{author}{\bibfnamefont{G.}~\bibnamefont{Kramposthuber}},
  \bibinfo{journal}{Europhys. Lett.} \textbf{\bibinfo{volume}{23}},
  \bibinfo{pages}{673} (\bibinfo{year}{1993}).

\bibitem[{\citenamefont{Yethiraj and {van Blaaderen}}(2003)}]{YethirajB03}
\bibinfo{author}{\bibfnamefont{A.}~\bibnamefont{Yethiraj}} \bibnamefont{and}
  \bibinfo{author}{\bibfnamefont{A.}~\bibnamefont{{van Blaaderen}}},
  \bibinfo{journal}{Nature} \textbf{\bibinfo{volume}{421}},
  \bibinfo{pages}{513} (\bibinfo{year}{2003}).

\bibitem[{\citenamefont{{L\"{o}wen} et~al.}(2005)\citenamefont{{L\"{o}wen},
  Esztermann, Wysocki, Allahyarov, Messina, Jusufi, Hoffmann, Gottwald, Kahl,
  Konieczny et~al.}}]{LowenETAL05}
\bibinfo{author}{\bibfnamefont{H.}~\bibnamefont{{L\"{o}wen}}},
  \bibinfo{author}{\bibfnamefont{A.}~\bibnamefont{Esztermann}},
  \bibinfo{author}{\bibfnamefont{A.}~\bibnamefont{Wysocki}},
  \bibinfo{author}{\bibfnamefont{E.}~\bibnamefont{Allahyarov}},
  \bibinfo{author}{\bibfnamefont{R.}~\bibnamefont{Messina}},
  \bibinfo{author}{\bibfnamefont{A.}~\bibnamefont{Jusufi}},
  \bibinfo{author}{\bibfnamefont{N.}~\bibnamefont{Hoffmann}},
  \bibinfo{author}{\bibfnamefont{D.}~\bibnamefont{Gottwald}},
  \bibinfo{author}{\bibfnamefont{G.}~\bibnamefont{Kahl}},
  \bibinfo{author}{\bibfnamefont{M.}~\bibnamefont{Konieczny}},
  \bibnamefont{et~al.}, \bibinfo{journal}{J. Phys.: Conf. Series}
  \textbf{\bibinfo{volume}{11}}, \bibinfo{pages}{207} (\bibinfo{year}{2005}).

\bibitem[{\citenamefont{{van Blaaderen} et~al.}(2013)\citenamefont{{van
  Blaaderen}, Dijkstra, {van Roij}, Imhof, Kamp, Kwaadgras, Vissers, and
  Liu}}]{vanBlaaderenETAL13}
\bibinfo{author}{\bibfnamefont{A.}~\bibnamefont{{van Blaaderen}}},
  \bibinfo{author}{\bibfnamefont{M.}~\bibnamefont{Dijkstra}},
  \bibinfo{author}{\bibfnamefont{R.}~\bibnamefont{{van Roij}}},
  \bibinfo{author}{\bibfnamefont{A.}~\bibnamefont{Imhof}},
  \bibinfo{author}{\bibfnamefont{M.}~\bibnamefont{Kamp}},
  \bibinfo{author}{\bibfnamefont{B.~W.} \bibnamefont{Kwaadgras}},
  \bibinfo{author}{\bibfnamefont{T.}~\bibnamefont{Vissers}}, \bibnamefont{and}
  \bibinfo{author}{\bibfnamefont{B.}~\bibnamefont{Liu}}, \bibinfo{journal}{Eur.
  Phys. J. Special Topics} \textbf{\bibinfo{volume}{222}},
  \bibinfo{pages}{2895} (\bibinfo{year}{2013}).

\bibitem[{\citenamefont{Smoukov et~al.}(2009)\citenamefont{Smoukov, Gangwal,
  Marquezbc, and Velev}}]{SmoukovGMV09}
\bibinfo{author}{\bibfnamefont{S.~K.} \bibnamefont{Smoukov}},
  \bibinfo{author}{\bibfnamefont{S.}~\bibnamefont{Gangwal}},
  \bibinfo{author}{\bibfnamefont{M.}~\bibnamefont{Marquezbc}},
  \bibnamefont{and} \bibinfo{author}{\bibfnamefont{O.~D.} \bibnamefont{Velev}},
  \bibinfo{journal}{Soft Matter} \textbf{\bibinfo{volume}{5}},
  \bibinfo{pages}{1285} (\bibinfo{year}{2009}).

\bibitem[{\citenamefont{Ivlev et~al.}(2012)\citenamefont{Ivlev, L\"{o}wen,
  Morfill, and Royall}}]{IvlevLMR12}
\bibinfo{author}{\bibfnamefont{A.}~\bibnamefont{Ivlev}},
  \bibinfo{author}{\bibfnamefont{H.}~\bibnamefont{L\"{o}wen}},
  \bibinfo{author}{\bibfnamefont{G.}~\bibnamefont{Morfill}}, \bibnamefont{and}
  \bibinfo{author}{\bibfnamefont{C.~P.} \bibnamefont{Royall}},
  \emph{\bibinfo{title}{Complex Plasmas and Colloidal Dispersions:
  Particle-Resolved Studies of Classical Liquids and Solids}}
  (\bibinfo{publisher}{World Scientific}, \bibinfo{address}{London},
  \bibinfo{year}{2012}).

\bibitem[{\citenamefont{Israelachvili}(2011)}]{Israelachvili11}
\bibinfo{author}{\bibfnamefont{J.~N.} \bibnamefont{Israelachvili}},
  \emph{\bibinfo{title}{Intermolecular and Surface Forces}}
  (\bibinfo{publisher}{Academic Press}, \bibinfo{address}{London},
  \bibinfo{year}{2011}).

\bibitem[{\citenamefont{Fejer et~al.}(2011)\citenamefont{Fejer, Chakrabarti,
  and Wales}}]{fejer_self-assembly_2011}
\bibinfo{author}{\bibfnamefont{S.~N.} \bibnamefont{Fejer}},
  \bibinfo{author}{\bibfnamefont{D.}~\bibnamefont{Chakrabarti}},
  \bibnamefont{and} \bibinfo{author}{\bibfnamefont{D.~J.} \bibnamefont{Wales}},
  \bibinfo{journal}{Soft Matter} \textbf{\bibinfo{volume}{7}},
  \bibinfo{pages}{3553} (\bibinfo{year}{2011}).

\bibitem[{\citenamefont{Paramonov and Yaliraki}(2005)}]{paramonov05}
\bibinfo{author}{\bibfnamefont{L.}~\bibnamefont{Paramonov}} \bibnamefont{and}
  \bibinfo{author}{\bibfnamefont{S.~N.} \bibnamefont{Yaliraki}},
  \bibinfo{journal}{\jcp} \textbf{\bibinfo{volume}{123}},
  \bibinfo{pages}{194111} (\bibinfo{year}{2005}).

\bibitem[{\citenamefont{Fejer et~al.}(2010)\citenamefont{Fejer, Chakrabarti,
  and Wales}}]{FejerCW10}
\bibinfo{author}{\bibfnamefont{S.~N.} \bibnamefont{Fejer}},
  \bibinfo{author}{\bibfnamefont{D.}~\bibnamefont{Chakrabarti}},
  \bibnamefont{and} \bibinfo{author}{\bibfnamefont{D.~J.} \bibnamefont{Wales}},
  \bibinfo{journal}{ACS Nano} \textbf{\bibinfo{volume}{4}},
  \bibinfo{pages}{219} (\bibinfo{year}{2010}).

\bibitem[{\citenamefont{Olesen et~al.}(2011)\citenamefont{Olesen, Fejer,
  Chakrabarti, and Wales}}]{olesen_2013}
\bibinfo{author}{\bibfnamefont{S.~W.} \bibnamefont{Olesen}},
  \bibinfo{author}{\bibfnamefont{S.~N.} \bibnamefont{Fejer}},
  \bibinfo{author}{\bibfnamefont{D.}~\bibnamefont{Chakrabarti}},
  \bibnamefont{and} \bibinfo{author}{\bibfnamefont{D.~J.} \bibnamefont{Wales}},
  \bibinfo{journal}{RSC Adv.} \textbf{\bibinfo{volume}{3}},
  \bibinfo{pages}{3553} (\bibinfo{year}{2011}).

\bibitem[{\citenamefont{Wales}(2005)}]{Wales05}
\bibinfo{author}{\bibfnamefont{D.~J.} \bibnamefont{Wales}},
  \bibinfo{journal}{Phil. Trans. Roy. Soc. A} \textbf{\bibinfo{volume}{363}},
  \bibinfo{pages}{357} (\bibinfo{year}{2005}).

\bibitem[{\citenamefont{Chakrabarti and Wales}(2009)}]{ChakrabartiW09}
\bibinfo{author}{\bibfnamefont{D.}~\bibnamefont{Chakrabarti}} \bibnamefont{and}
  \bibinfo{author}{\bibfnamefont{D.~J.} \bibnamefont{Wales}},
  \bibinfo{journal}{Phys. Chem. Chem. Phys.} \textbf{\bibinfo{volume}{11}},
  \bibinfo{pages}{1970} (\bibinfo{year}{2009}).

\bibitem[{\citenamefont{Boerdijk}(1952)}]{boerdijk52}
\bibinfo{author}{\bibfnamefont{A.~H.} \bibnamefont{Boerdijk}},
  \bibinfo{journal}{Philips Res. Rep.} \textbf{\bibinfo{volume}{7}},
  \bibinfo{pages}{303} (\bibinfo{year}{1952}).

\bibitem[{\citenamefont{Erickson}(1973)}]{tubularpacking73}
\bibinfo{author}{\bibfnamefont{R.~O.} \bibnamefont{Erickson}},
  \bibinfo{journal}{Science} \textbf{\bibinfo{volume}{181}},
  \bibinfo{pages}{705} (\bibinfo{year}{1973}).

\bibitem[{\citenamefont{Chen et~al.}(2011{\natexlab{b}})\citenamefont{Chen,
  Diesel, Whitmer, Bae, Luijten, and Granick}}]{Chen2011}
\bibinfo{author}{\bibfnamefont{Q.}~\bibnamefont{Chen}},
  \bibinfo{author}{\bibfnamefont{E.}~\bibnamefont{Diesel}},
  \bibinfo{author}{\bibfnamefont{J.~K.} \bibnamefont{Whitmer}},
  \bibinfo{author}{\bibfnamefont{S.~C.} \bibnamefont{Bae}},
  \bibinfo{author}{\bibfnamefont{E.}~\bibnamefont{Luijten}}, \bibnamefont{and}
  \bibinfo{author}{\bibfnamefont{S.}~\bibnamefont{Granick}},
  \bibinfo{journal}{Journal of the American Chemical Society}
  \textbf{\bibinfo{volume}{133}}, \bibinfo{pages}{7725}
  (\bibinfo{year}{2011}{\natexlab{b}}).

\bibitem[{\citenamefont{Gay and Berne}(1981)}]{gayberne}
\bibinfo{author}{\bibfnamefont{J.~G.} \bibnamefont{Gay}} \bibnamefont{and}
  \bibinfo{author}{\bibfnamefont{B.~J.} \bibnamefont{Berne}},
  \bibinfo{journal}{\jcp} \textbf{\bibinfo{volume}{74}}, \bibinfo{pages}{3306}
  (\bibinfo{year}{1981}).

\bibitem[{\citenamefont{Iwashita and Kimura}(2013)}]{Iwashita2013}
\bibinfo{author}{\bibfnamefont{Y.}~\bibnamefont{Iwashita}} \bibnamefont{and}
  \bibinfo{author}{\bibfnamefont{Y.}~\bibnamefont{Kimura}},
  \bibinfo{journal}{Soft Matter} \textbf{\bibinfo{volume}{9}},
  \bibinfo{pages}{10694} (\bibinfo{year}{2013}).

\bibitem[{\citenamefont{Preisler et~al.}(2013)\citenamefont{Preisler, Vissers,
  Smallenburg, Munaò, and Sciortino}}]{Preisler2013}
\bibinfo{author}{\bibfnamefont{Z.}~\bibnamefont{Preisler}},
  \bibinfo{author}{\bibfnamefont{T.}~\bibnamefont{Vissers}},
  \bibinfo{author}{\bibfnamefont{F.}~\bibnamefont{Smallenburg}},
  \bibinfo{author}{\bibfnamefont{G.}~\bibnamefont{Munaò}}, \bibnamefont{and}
  \bibinfo{author}{\bibfnamefont{F.}~\bibnamefont{Sciortino}},
  \bibinfo{journal}{J. Phys. Chem. B} \textbf{\bibinfo{volume}{117}},
  \bibinfo{pages}{9540} (\bibinfo{year}{2013}).

\bibitem[{\citenamefont{Doye and Wales}(1999)}]{DoyeW99}
\bibinfo{author}{\bibfnamefont{J.~P.~K.} \bibnamefont{Doye}} \bibnamefont{and}
  \bibinfo{author}{\bibfnamefont{D.~J.} \bibnamefont{Wales}},
  \bibinfo{journal}{Phys. Rev. B} \textbf{\bibinfo{volume}{59}},
  \bibinfo{pages}{2292} (\bibinfo{year}{1999}).

\end{thebibliography}
\end{document}